\RequirePackage{silence}
\documentclass[amsmath,amssymb,aps,floatfix,twocolumn, aas_macros]{revtex4-2}
\usepackage[utf8]{inputenc}
\usepackage[colorlinks=true,citecolor=blue,allcolors=blue]{hyperref}
\usepackage{amsmath,amsthm,amsfonts,amssymb,amscd}
\usepackage{graphicx}
\usepackage{lastpage}
\usepackage{enumerate}
\usepackage{mathrsfs}
\usepackage{MnSymbol}
\usepackage{bm}
\usepackage{braket}
\usepackage{tabularx}
\usepackage{makecell}
\usepackage[version=4]{mhchem}
\usepackage{natbib}

\newcommand\mati{\begin{matrix}}
\newcommand\matf{\end{matrix}}
\newcommand\bmati{\begin{bmatrix}}
\newcommand\bmatf{\end{bmatrix}}
\newcommand\pmati{\begin{pmatrix}}
\newcommand\pmatf{\end{pmatrix}}

\newcommand{\pmat}[1]{\pmati #1 \pmatf}

\newcommand{\abs}[1]{\left\lvert#1\right\rvert}

\newcommand{\quotes}[1]{``#1''}
\newcommand{\wignerjjj}[6]{\pmat{ #1 & #2 & #3 \\ #4 & #5 & #6 }}
\newcommand{\vecp}[1]{\vec{#1}\mkern2mu\vphantom{#1}'}
\newcommand{\hatp}[1]{\hat{#1}\mkern2mu\vphantom{#1}'}
\newcommand{\ldotstwo}{\mathinner{{\ldotp}{\ldotp}}}

\begin{document}

\author{ Joshua Forer }

\affiliation{Columbia Astrophysics Laboratory, Columbia University \\ New York, New York 10027, USA}

\title{
  Rotational excitation of asymmetric-top molecular ions by electron impact: application to \ce{H2O+}, \ce{HDO+}, and \ce{D2O+}
}

\date{\today}
\begin{abstract}
  The rotational excitation of the three asymmetric-top molecular ion isotopologues \ce{H2O+}, \ce{HDO+}, and \ce{D2O+} is studied theoretically using a combined framework of electron-molecule R-matrix scattering theory, multichannel quantum-defect theory, frame transformation theory, and the Coulomb-Born approximation.
  The latter two have been adapted here for asymmetric-top rotors.
  State-resolved cross sections and kinetic rate coefficients for transitions from the rotational ground state of the ions are presented.
  State-resolved rate coefficients for all calculated transitions $N=0\ldotstwo4$ are included as supplementary material and will be made available through the EMAA database.
\end{abstract}

\maketitle


\section{Introduction}

Collisions between electrons and molecules provide an important energy redistribution channel in molecular gases, such as interstellar clouds, planetary atmospheres, and various plasma environments.
Electrons can lose their energy by exciting any combination of a molecule's degrees of freedom, but can also gain energy via de-excitation --- often leading to a cooling process that, depending on the density of electrons and molecules, can proceed much faster than radiative decay.
It is therefore critical to know the state-to-state cross sections or rate coefficients for these processes if one wishes to infer a variety of observables that depend on the internal state distribution of a population of molecules.

Ignoring spin, slow electrons typically exchange energy with the rotational motion of a molecule, given that the excitation thresholds are most often smaller and denser for rotational excitation than they are for vibrational or electronic excitation.
This type of collision is particularly important in sparse and cold media, where the molecules have time to cool to their ground vibrational and electronic state and the electrons do not have the energy to excite the molecule vibrationally or electronically.
This includes ground-based experiments \cite{znotins2025electron}, but also a variety of space environments, e.g.,
cometary comae,
planetary nebulae,
interstellar clouds.
Positive molecular ions in particular play a critical role in the early chemistry of interstellar clouds when they are in their most diffuse phase.
Over millions of years, a cloud can collapse under its own gravity and form stars, which largely determine galactic structure \cite{gerin2016interstellar}.
This collapse can be impeded by a cloud's ionization fraction, which can couple it to the local magnetic field and completely prevent its collapse.
Diffuse clouds typically have kinetic temperatures between 40 and 130~K \cite{shull2021fuv}, where neutral-neutral reactions tend to be much slower than barrierless ion-electron and ion-neutral reactions.
Therefore, several ions help trace important diffuse cloud parameters, such as the molecular fraction and cosmic ray ionization rate of atomic hydrogen\cite{indriolo2015herschel}.
To make such inferences based on observed column densities, it is necessary to understand ion chemistry in these environments.

In the context of rotation, molecules are typically classified by their principal moments of inertia along the three orthogonal molecule-fixed axes $\{A,B,C\}$, such that the principal moments of inertia satisfy $I_A \le I_B \le I_C$.
The broadest category of rotors is the \textit{asymmetric top}, which has three different, nonzero principal moments of inertia.
While many important astrochemical molecules are linear (e.g., CO: $I_A=0,I_B=I_C$) or symmetric tops (e.g., \ce{H3+}: $0<I_A=I_B<I_C$), most molecules fall under the category of the asymmetric top (e.g., \ce{H2O} and \ce{H2O+}).
However, there are few studies on electron-impact rotational excitation of such molecules, and this is especially true for asymmetric top molecular ions.
With the advent of the powerful new observational capabilities of, e.g., the James Webb Space Telescope, the next-generation Very Large Array, and the Atacama Large Millimeter/submillimeter Array Wideband Sensitivity Upgrade, it is reasonable to anticipate the detection of new molecules in space and new problems that require accurate knowledge of state-to-state chemistry for a wide range of molecules.

This work presents a theoretical approach to determining the electron-impact rotational (de-)excitation cross sections for asymmetric-top molecular ions, with or without a significant electric dipole moment, by combining accurate \textit{ab initio} R-matrix electron-molecule scattering calculations, multichannel quantum-defect theory (MQDT), frame transformation theory, and the Coulomb-Born approximation, applied to the \ce{H2O+}, \ce{HDO+}, and \ce{D2O+} molecular ion isotopologues.
Section \ref{sec:abinitio} discusses the \textit{ab initio} R-matrix scattering calculations, implementation of the rotational frame transformation for asymmetric-top rotors, and application of MQDT to obtain rotationally resolved integrated cross sections.
Next, Section \ref{sec:bornclosure} introduces the Coulomb-Born approximation, which provides a perturbative asymptotic correction to dipolar rotational (de-)excitation transitions.
Section \ref{sec:results} presents the results obtained from applying these developments to the electron-impact (de-)excitation of \ce{H2O+}, \ce{HDO+}, and \ce{D2O+}.
Concluding remarks and a brief discussion on the application of this method to neutral molecules are given in Section \ref{sec:conclusion}.

\section{Electron Scattering}
\label{sec:abinitio}

\subsection{R-matrix scattering}
\label{sec:rmat}
In typical \textit{ab initio} electron-scattering calculations, the scattering electron is often represented as a wave that is decomposed into an infinite series of partial waves, each determined by its orbital angular momentum $\vec{l}$ and the projection $\lambda$ of $\vec{l}$ on the molecule-fixed $\hat{z}$-axis.
This series is typically truncated to the lowest partial waves that carry the most information about the short-range physics due to computational restrictions.
These scattering calculations are performed at the equilibrium geometry of each target ion using the \texttt{UKRmol+ v3.2.0} software suite \cite{masin2020ukrmolx,houfek2024scripts,ukrmol-in_3_2}; see those works for more details on the method and its implementation.
The R-matrix method involves a division of the physical space of the scattering problem into two complementary regions: an inner and an outer region, defined by the R-matrix bounding sphere.
In this study, the bounding sphere is centered on the target molecule's center of mass with a radius of 13 bohr.
The inner and outer region solutions are matched at the R-matrix sphere, and the matching condition determines the R-matrix, used to obtain the S-matrix, which directly links observable scattering cross sections to the scattering calculations.

For each calculation, the target ion's wavefunction was described using a complete active space configuration interaction (CAS-CI) of Hartree-Fock molecular orbitals that were obtained from a self-consistent field (SCF) \texttt{psi4}\cite{psi4} calculation.
The partial-wave basis includes all partial waves $l=0\ldotstwo5$.
The lowest three electronic states at equilibrium are included in the R-matrix scattering calculations for all target ions.
The active space and electronic states used in the calculation are shown in Table \ref{tab:target}.

\begin{table}[]
  \centering
  \caption{
    Information on the \ce{H2O+}, \ce{HDO+}, and \ce{D2O+} targets is given here:
    their point groups at equilibrium,
    the electronic states used in the R-matrix calculations,
    the restricted, active, and virtual orbitals used in the R-matrix calculations,
    their $A$- and $B$-components of their electric dipole moment,
    their $A$, $B$, and $C$ rotational constants,
    and the five centrifugal distortion parameters used in the Watson A-reduced effective Hamiltonian (\ref{eqn:app:HCD4}).
    Electric dipole ($\mu_A,\mu_B$) units are in Debye; rotational constants and centrifugal distortion parameters are all given in cm$^{-1}$.
  }
  \begin{tabularx}{\textwidth}{c|c|c|c}
    \cline{1-4}
    molecule                             & \ce{H2O+}                                                          & \ce{HDO+}                                  & \ce{D2O+}                              \\ \cline{1-4}
    \makecell{point                     \\ group}                                                             & $C_{2v}$                                   & $C_{s}$                                 & $C_{2v}$                     \\ \cline{1-4}
    \makecell{electronic                \\ states}                                                            & \makecell{X$^2B_1$, A$^2A_1$,             \\ B$^2B_2$ }                              & \makecell{X$^2A''$, A$^2A'$, \\ B$^2A'$}                    & \makecell{X$^2B_1$, A$^2A_1$, \\ B$^2B_2$} \\ \cline{1-4}
    \makecell{restricted                \\ orbitals}                                                          & $1a_1$                                     & $1a'$                                   & $1a_1$                       \\ \cline{1-4}
    \makecell{active                    \\ orbitals}                                                          & \makecell{$2-4a_1$, $1b_1$,               \\ $1-2b_2$}                               & $2-5a'$, $1a''$               & \makecell{$2-4a_1$, $1b_1$, \\ $1-2b_2$}                     \\ \cline{1-4}
    \makecell{virtual                   \\ orbitals}                                                          & \makecell{$5-7a_1$, $2b_1$,               \\ $3-4b_2$}                               & $7-11a'$, $2a''$              & \makecell{$5-7a_1$, $2b_1$, \\ $3-4b_2$}                     \\ \cline{1-4}
    $\mu_A$ \hfill                       &                                                                    & 0.495 \hfill\cite{lauzin2015threshold}     &                                        \\
    $\mu_B$ \hfill                       & 2.398    \hfill\cite{endres2016cdms,muller2016ohx_h2ox_absorption} & -2.208 \hfill\cite{lauzin2015threshold}    & 2.135    \hfill\cite{duan2008infrared} \\
    $A$ \hfill                           & 29.0359  \hfill\cite{endres2016cdms,muller2016ohx_h2ox_absorption} & 23.5   \hfill\cite{lauzin2015threshold}    & 16.05622 \hfill\cite{duan2008infrared} \\
    $B$ \hfill                           & 12.42298 \hfill\cite{endres2016cdms,muller2016ohx_h2ox_absorption} & 7.8    \hfill\cite{lauzin2015threshold}    & 6.234696 \hfill\cite{duan2008infrared} \\
    $C$  \hfill                          & 8.46921  \hfill\cite{muller2016ohx_h2ox_absorption}                & 6.0      \hfill\cite{weis1989theoretical}  & 4.408003 \hfill\cite{duan2008infrared} \\
    $\Delta_N \times 10^{-3}$ \hfill     & 0.9893   \hfill\cite{muller2016ohx_h2ox_absorption}                & 0.311464 \hfill\cite{weis1989theoretical}  & 0.22708  \hfill\cite{duan2008infrared} \\
    $\Delta_{NK} \times 10^{-3}$ \hfill  & -5.18023 \hfill\cite{muller2016ohx_h2ox_absorption}                & 0.504    \hfill\cite{weis1989theoretical}  & -1.2719  \hfill\cite{duan2008infrared} \\
    $\Delta_{K} \times 10^{-3}$  \hfill  & 45.875   \hfill\cite{muller2016ohx_h2ox_absorption}                & 15.43    \hfill\cite{weis1989theoretical}  & 12.650   \hfill\cite{duan2008infrared} \\
    $\delta_{N} \times 10^{-3}$ \hfill   & 0.37846  \hfill\cite{muller2016ohx_h2ox_absorption}                & 0.097   \hfill\cite{weis1989theoretical}   & 0.07776  \hfill\cite{duan2008infrared} \\
    $\delta_{K} \times 10^{-3}$ \hfill   & 1.728    \hfill\cite{muller2016ohx_h2ox_absorption}                & 1.27     \hfill\cite{weis1989theoretical}  & 0.4558   \hfill\cite{duan2008infrared} \\
  \end{tabularx}
  \label{tab:target}
\end{table}

The R- and S-matrices are expressed in a basis of continuum electronic channels $\ket{nl\lambda}$, where $n$ enumerates the electronic state of the target and the set $\{l,\lambda\}$ determines the partial wave of the scattering electron.
Each ionic state hosts an infinite series of Rydberg states, which can have a significant effect on electron scattering, especially in the case of low-lying electronic states.
At the equilibrium geometry, the first excited state is about 2~eV above the ground state, while the second state sits somewhat higher, around 5.6~eV above the ground state.
The R-matrices are extracted from the scattering calculations, and all channels connected to the electronically excited states ($n>1$) are eliminated via an MQDT closed-channel elimination procedure to obtain the so-called physical R-matrix, which is expressed in a basis of ground-state electronic channels that include the effects of the closed excited-state electronic channels.
Then, the R-matrices are transformed into the K-matrices, as described by \citet{hvizdos2023bound}.

\subsection{Rotational Frame Transformation}
\label{sec:RFT}

There are multiple physically equivalent matrices that can be obtained from electron-scattering calculations, and they are related by several transformations, e.g., the K-matrix, the T-matrix, and the S-matrix.
Cross sections can be formulated in terms of the S-matrix, which is obtained from the K-matrix via a Cayley transform,
\begin{equation}
  S = (I + iK)(I-iK)^{-1},
  \label{eqn:cayley}
\end{equation}
where $I$ is the identity matrix, $K$ is the real and symmetric K-matrix, and $S$ is the complex-symmetric S-matrix.
Next, the $S$-matrix is transformed from a basis of partial waves represented by real-valued spherical harmonics $X_{l\lambda}$ (used in the quantum chemistry calculations) to a basis of partial waves represented by complex-valued spherical harmonics $Y_l^\lambda$.
Given the standard definition of $X_{l\lambda}$ in terms of $Y_l^\lambda$, this inverse transformation is given by
\begin{equation}
  Y_l^\lambda =
  \begin{cases}
    \frac{(-1)^\lambda}{\sqrt{2}} \left( X_{l\abs{\lambda}} + iX_{l-\abs{\lambda}}  \right) & \lambda > 0, \\
    X_{l0} & \lambda = 0, \\
    \frac{1}{\sqrt{2}} \left( X_{l\abs{\lambda}} -i X_{l-\abs{\lambda}} \right) & \lambda < 0.
  \end{cases}
\end{equation}
This transformation is necessary here because the rest of this approach is formulated in terms of the complex-valued spherical harmonics.

Once the S-matrix is in the basis of complex-valued spherical harmonics, the rotational frame transformation can take place using the rotational wavefunctions of the target.
The rotational motion of an isolated molecule in the absence of external fields can be described, to the lowest order, by the rigid rotor approximation, where the nuclei are assumed to rotate about fixed equilibrium positions while the molecular geometry remains unchanged.
The asymmetric-top wavefunctions can be expressed in the basis of symmetric-top wavefunctions,
\begin{equation}
  \begin{aligned}
    \ket{N\tau M}
    &= \sum\limits_K c^{(N\tau)}_K \ket{NKM}
    \\
    &=  \sqrt{\frac{2N+1}{8\pi^2}} \sum\limits_K c^{(N\tau)}_K D^N_{MK}(\alpha, \beta, \gamma)^*,
  \end{aligned}
  \label{eqn:wf_rot}
\end{equation}
where $\hat{\boldsymbol N}$ is the angular momentum of the nuclei, $K$ and $M$ are the projections of $\hat{\boldsymbol N}$ on the molecule-fixed and space-fixed $\hat{z}$-axes (respectively), $D^J_{MK}(\alpha,\beta,\gamma)$ is the Wigner D-matrix as a function of the Euler angles ($\alpha,\beta,\gamma$), $c^{(N\tau)}_K$ is a mixing coefficient that is determined by numerical diagonalization of (\ref{eqn:app:HRR}--\ref{eqn:app:HCD4}), and $\tau$ enumerates the $2N+1$ asymmetric-top rotational states for each value of $N$.
Each rotational state can be labelled with the \textit{approximate} quantum numbers $K_a$ and $K_c$, which represent the absolute magnitude of the dominant projection $K$ on the molecule-fixed $A$ and $C$ axes, respectively.
Each $\tau$ effectively enumerates different combinations of the labels $K_a$ and  $K_c$.

For molecules with identical nuclei like \ce{H2O+} and \ce{D2O+} with a $C_2$ symmetry axis that coincides with the $B$-axis, the permutation symmetry of the nuclei separates the rotational levels into two exchange-symmetry classes that are distinguished by the parity of $K_a+K_c$.
For \ce{H2O+}, even and odd values of $K_a+K_c$ correspond, respectively, to \textit{para} and \textit{ortho} nuclear-spin symmetry species.
However, this ortho/para identification reverses for \ce{D2O+}.
Spin-agnostic electron collisions, similar to radiative transitions, preserve this symmetry.
Lacking any identical nuclei, \ce{HDO+} does not have separate ortho/para species.

While the rotational frame transformation can be carried out on any of the aforementioned equivalent matrices; it is applied here to the S-matrix, which is still expressed in a basis of asymptotic body-frame electronic channels.
To obtain the S-matrix in a basis of asymptotic laboratory-frame rotational channels $\ket{N\tau nl}$, the rotational frame transformation
\begin{equation}
  \hspace{-.5em}
  \begin{aligned}
    &S^{J}_{N\tau nl,N'\tau'n'l'} = \frac{\sqrt{(2N+1)(2N'+1)}}{2J+1} \\
    & \times
    \sum\limits_{\substack{\Omega\\\lambda\lambda'}}
    \sum\limits_{KK'} c^{(N\tau)*}_{K} c^{(N'\tau')}_{K'} C^{J\Omega}_{NK,l\lambda} C^{J\Omega}_{N'K',l'\lambda'}
    S_{nl\lambda,n'l'\lambda'},
  \end{aligned}
  \hfill
  \label{eqn:RFT}
\end{equation}
is carried out, where $N$ is the magnitude of the rotational angular momentum of the target, $J$ is the magnitude of the total angular momentum $\vec{J}=\vec{N}+\vec{l}$, $c^{(N\tau)}_K$ are the rotational wavefunction expansion coefficients (\ref{eqn:wf_rot}), $\Omega=K+\lambda$, $C^{J\Omega}_{NK,l\lambda}$ are Clebsch-Gordan coefficients, $S_{nl\lambda,n'l'\lambda'}$ is the electronic S-matrix element relating the asymptotic electronic channels $\ket{nl\lambda}$ and $\ket{n'l'\lambda'}$, and $S^J_{N\tau nl,N'\tau'n'l'}$ is the S-matrix element relating the asymptotic \textit{roelectronic} (rotational and electronic) channels $\ket{N\tau nl}$ and $\ket{N'\tau'n'l'}$.
The quantity $\Omega=K+\lambda$ can be thought of as an intermediate coupling label, rather than as an actual projection on the molecular axis.
The rotational frame transformation is applied separately for each $J$, resulting in a block-diagonal rotational S-matrix with respect to $J$ because it is a conserved quantity during the collision.

Ortho/para symmetry is automatically enforced (numerically) in this work because the electron-scattering calculations are performed in the full $C_{2v}$ point group of \ce{H2O+} and \ce{D2O+} with the $C_2$ axis aligned with the $\hat{z}$-axis.
If the electron-scattering calculations are carried out in a reduced symmetry, this will likely not hold.
In such a case, one may need to expand the symmetry of the scattering calculations to the natural point group (e.g., with correlation tables) of the target molecule.

\subsection{Cross Sections}

At this point, the roelectronic S-matrix (\ref{eqn:RFT}) is expressed in a basis of channels that are not necessarily open at all scattering energies.
The closed channels produce resonances in the open channels that can significantly affect the resulting cross sections.
This effect is captured by the MQDT closed-channel elimination procedure which proceeds as follows.
First, the rotational S-matrix (\ref{eqn:RFT}) is partitioned into blocks corresponding to open ($o$) and closed ($c$) channels,
\begin{equation}
  S^{J} = \pmat{S_{oo}&S_{oc}\\S_{co}&S_{cc}}, \;
  \beta_{ij}(E_\text{tot}) = \frac{\pi}{\sqrt{2(E_i - E_\text{tot})}} \delta_{ij},
  \label{eqn:Spartition}
\end{equation}
where $i$ and $j$ enumerate \textit{closed} scattering channels, $\beta(E_\text{tot})$ is a diagonal matrix in the basis of closed channels,
$E_\text{tot}$ is the total scattering energy, $E_i$ is the threshold energy for channel $i$, and $\delta_{ij}$ is the Kronecker delta.
Then, the \textit{physical} S-matrix, expressed in a basis of energetically available channels, is obtained by the matrix equation
\begin{equation}
  S^{J,\text{phys}}(E_\text{tot}) = S_{oo}
  - S_{oc} \left[S_{cc} - e^{-2i\beta(E_\text{tot})}\right]^{-1} S_{co},
  \label{eqn:CCEP}
\end{equation}
which immediately leads to the rotational (de-)excitation cross sections,
\begin{equation}
  \begin{aligned}
    &\sigma_{N\tau n\to N'\tau'n}(E_\text{el}) =
    \frac{\pi}{2m_\text{e}E_\text{el}}
    \sum\limits_J \frac{2J+1}{2N+1} \\
    &\qquad \times \sum\limits_{ll'}^\infty \abs{S^{J,\text{phys}}_{N'\tau'nl',N\tau nl}(E_\text{tot}) - \delta_{N'\tau'nl',N\tau nl}}^2
  \end{aligned}
  \label{eqn:xs_mqdt}
\end{equation}
where $E_\text{el}=E_\text{tot}-E_i$ is the scattering electron energy from the initial state $i$, $\delta$ is the Kronecker delta, and $m_\text{e}$ is the electron mass.

The electronic state $n$ is the same for all rotational states considered in this study, hence the use of $n$ instead of $n'$ in the notation for the final state of the cross section.
The sum over $l,l'$ is taken over all available values in the partial wave basis $l=0\ldotstwo5$.
An important note here is that the S-matrix is taken to be constant in energy.
In this work, it is evaluated at 0.25~eV above the ground electronic state.
At and below this energy, the scattering phases are smooth and all nearly constant at the ions' equilibrium position.

\section{Born Closure}
\label{sec:bornclosure}

The previous section describes how to obtain cross sections from \textit{ab initio} scattering calculations, and is generally applicable to any method that can produce body-frame matrices, such as the K- or S-matrices.
However, these calculations are limited by the partial-wave basis used to represent the scattering electron.
This can result in a significant underestimation of the rotational (de-)excitation cross section for molecules (ions and neutrals) with a significant electric dipole moment because it couples partial waves that differ by unity; $l=0$ couples to $l=1$, which couples to $l=2$, and so on.
Given a strong electric dipole moment, this coupling can extend to much higher partial waves than are included in the scattering basis.
These higher partial waves are included in a perturbative approach known as the Coulomb-Born approximation, originally introduced by \citet{boikova1968rotational}, and has been applied to linear \cite{boikova1968rotational,chu1974rotational,rabadan1998rotions,faure2001electron,faure2002electron,faure2003rate,hamilton2016electron,forer2024kinetic}, symmetric top molecules \cite{chu1975rotational,faure2002electron}, and very recently to the asymmetric-top ion \ce{D2H+} \cite{znotins2025electron}.
The main aspects of the Coulomb-Born approximation for asymmetric top ions are presented here for completeness.

\subsection{The Coulomb-Born Approximation}

Consider the potential energy for the electron-ion ($Z=1$) interaction at large distances, expanded into a spherical harmonics basis in the molecular frame (MF), \cite{stogryn1966molecular,itikawa1971electron,chu1975rotational}
\begin{equation}
  \begin{aligned}
    V &= -\frac{1}{r} - H'_\text{MF}\\
     &\equiv
    -\frac{1}{r}
    - \sum\limits_{\xi=1}^\infty
    \sum\limits_{\mu=-\xi}^\xi
    \sqrt{\frac{4\pi}{2\xi+1}}
    \frac{Q_{\xi\mu}}{r^{\xi+1}}
    Y_\xi^\mu(\chi,\psi),
    \label{eqn:CB_HMF}
  \end{aligned}
\end{equation}
where $(\chi,\psi)$ are angles in the molecular frame, $Y_\xi^\mu(\chi,\psi)$ are complex-valued spherical harmonics, $r$ is the distance between the scattering electron and the molecule's center of mass, $H'_\text{MF}$ is the MF perturbation to the Coulomb potential, $\xi$ is the order of each term in the multipole expansion, and $Q_{\xi\mu}$ are the components of the spherical multipole tensor of rank $\xi$,
\begin{equation}
  \begin{aligned}
    Q_{\xi\mu} &\equiv
    \frac{\xi!}{\sqrt{(\xi+\mu)!(\xi-\mu)!}} (-1)^\mu \\
    &\quad \times \sum\limits_{j=0}^\mu (-i)^{\mu-j} \pmat{\mu\\ j}
    M^{\xi}_{x^jy^{\mu-j}z^{\xi-\mu}},
  \end{aligned}
  \label{eqn:potv_multipoles}
\end{equation}
which are given in terms of the Cartesian multipole moment tensor,
\begin{equation}
  \begin{aligned}
    M^\xi_{x^jy^{\mu-j}z^{\xi-\mu}} &= \frac{(-1)^\xi}{\xi!}
    \int d\vecp{r} \rho(\vecp{r}) {r'}^{2\xi+1} \\
    &\quad \times \frac{\partial^\xi}{\partial_{x'}^j \partial_{y'}^{\mu-j} \partial_{z'}^{\xi-\mu}} \left(\frac{1}{r'}\right),
  \end{aligned}
  \label{eqn:multipoles}
\end{equation}
which depends on the target's charge density $\rho(\vecp{r})$.
The integrated Coulomb-Born rotational (de-)excitation cross section can be expressed as
\begin{subequations}
  \begin{align}
    &\sigma_{N\tau\to N'\tau'} =
    \frac{16\pi}{2\xi+1}
    \frac{k'}{k}
    (2N'+1) \sum\limits_\xi  \nonumber\\
    &\qquad \times
    \sum\limits_{ll'} \wignerjjj{l}{l'}{\xi}{0}{0}{0}^2 (2l+1)(2l'+1) \abs{M^\xi_{ll'}}^2
    \label{eqn:CB_xs_electronic} \\
    &\qquad \times
    \abs{\sum\limits_\mu Q_{\xi\mu} G^{\xi\mu}_{N\tau,N'\tau'}}^2
    \label{eqn:CB_xs_angular}
  \end{align}
  \label{eqn:CB_xs_solved}
\end{subequations}
which cleanly separates into the electronic term (\ref{eqn:CB_xs_electronic}), given in terms of the Coulomb integrals
\begin{equation}
  M^\xi_{ll'} = \frac{1}{kk'} \int\limits_0^\infty
  dr F_l(kr) r^{-\xi-1} F_{l'}(k'r),
  \label{eqn:CB_M}
\end{equation}
and an angular part (\ref{eqn:CB_xs_angular}), given in terms of the spherical multipole tensor and the quantity
\begin{equation}
  G^{\xi\mu}_{N\tau,N'\tau'} =
  \sum\limits_{KK'}
  (-1)^K
  c^{(N\tau)}_K c^{(N'\tau')*}_{K'}
  \wignerjjj{N}{N'}{\xi}{-K}{K'}{-\mu}.
  \label{eqn:CB_G}
\end{equation}
Even though the coefficients $c^{(N\tau)}_K$ originally come from the diagonalization of a real and symmetric matrix, the eigenvectors may be rotated and become complex, in which case the conjugation matters.
Such a case arises if the rotational Hamiltonian (\ref{eqn:app:HRR}, \ref{eqn:app:HCD4}) is diagonalized in one coordinate system, but the scattering calculations are performed in another.

\subsection{Dipolar Transitions}

Considering only the dipole term ($\xi=1$) of the multipole expansion (\ref{eqn:CB_HMF}), the angular term (\ref{eqn:CB_xs_angular}) is directly proportional to the Einstein coefficients for dipolar radiative decay,
\begin{equation}
  A_{N\tau\leftarrow  N'\tau'} =
  (2N+1)
  \frac{4\omega^3}{3c^3} \abs{\sum\limits_\mu Q_{1\mu} G^{1\mu}_{N\tau,N'\tau'}}^2,
  \label{eqn:EinstA}
\end{equation}
which is given in atomic units, and the electronic term (\ref{eqn:CB_xs_electronic}) can now be solved analytically.
The cross section (\ref{eqn:CB_xs_solved}) then becomes \cite{znotins2025electron}
\begin{equation}
  \begin{gathered}
    \sigma_{N\tau\to N'\tau'} =
    4\pi
    \frac{k'}{k}
    \frac{c^3}{\omega^3}
    \frac{2N'+1}{2N+1}
    A_{N\tau\leftarrow  N'\tau'} f^{l_\text{max}}, \\
    f^{l_\text{max}} = \sum\limits_{ll'}^{l_\text{max}} \wignerjjj{l}{l'}{1}{0}{0}{0}^2 (2l+1)(2l'+1) \abs{M^1_{ll'}}^2,
  \end{gathered}
  \label{eqn:CB_xs_EinstA}
\end{equation}
where $l_\text{max}$ can be taken to be finite or infinite.
The dipole selection rules in terms of rotational levels for asymmetric tops are given in Table \ref{tab:selection}.
\begin{table}[]
  \centering
  \caption{
    Allowed dipole selection rules for electric dipoles along the $A$, $B$, and $C$ molecule-fixed inertial axes.
    Dipolar transitions for molecules with components along multiple axes must satisfy at least one of the selection rules.
  }
  \centering
  \begin{tabular}{c|c|c|c}
    \cline{1-4}
    \makecell{Dipole\\Component} & $\Delta N$ & $\Delta K_a$ & $\Delta K_c$ \\\cline{1-4}
    $\mu_A$          & 0 ($N\ne0$), $\pm1$ & 0 & $\pm1$ \\
    $\mu_B$          & 0 ($N\ne0$), $\pm1$ & $\pm1$ & $\pm1$ \\
    $\mu_C$          & 0 ($N\ne0$), $\pm1$ & $\pm1$ & 0 \\\cline{1-4}
  \end{tabular}
  \label{tab:selection}
\end{table}
The integral $M^1_{ll'}$ only contributes to the cross section (\ref{eqn:CB_xs_EinstA}) when $l'=l\pm1$ because of the degenerate triangle inequality imposed by the Wigner 3-$j$ symbol relating each of $\{l,l',1\}$, which allows one to make use of the analytic expression of $M^1_{l,l+1}$ \cite{alder1956study},
\begin{equation}
  \begin{aligned}
    &M^1_{l,l+1} =
    -\frac{1}{2} \left(\frac{\eta'-\eta}{\eta'+\eta}\right)^{i(\eta'+\eta)}
    e^{-\pi\zeta/2} \frac{(-\chi_0)^{l+1}}{(2l+2)!} \\
    &\quad\times \abs{\Gamma(l+1+i\eta)} \abs{\Gamma(l+2+i\eta')} \\
    &\quad \times \Big[
      \eta \;{}_2F_1(l+1-i\eta,l+1-i\eta';2l+2;\chi_0) \\
      &\qquad -\eta' \frac{\abs{l+1+i\eta}^2}{(2l+2)(2l+3)} (-\chi_0) \\
      &\qquad \times  {}_2F_1(l+2-i\eta, l+2-i\eta';2l+4;\chi_0)
    \Big],
  \end{aligned}
  \label{eqn:CB_dip_M}
\end{equation}
where
\begin{equation}
  \begin{aligned}
    &\eta = -Z/k,&
    &\eta' = -Z/k',& \\
    &\zeta = \eta'-\eta,& \quad
    &\chi_0 = -4\eta\eta'/\zeta^2,&
  \end{aligned}
  \label{eqn:CB_dip_M_terms}
\end{equation}
and the functions $_2F_1(a,b;c;z)$ are the Gauss hypergeometric functions, defined in Appendix \ref{app:CB}.
Note that the expression $M^1_{l,l+1}$ (\ref{eqn:CB_dip_M}) corresponds to the quantity $-(M^{-2}_{l+1,l})/4$ in the work of \citet{alder1956study}.
When $l_\text{max}$ is taken to be infinite in (\ref{eqn:CB_xs_EinstA}), $f^{l_\text{max}}$ takes the form \cite{chu1974rotational,chu1975rotational}
\begin{equation}
  \begin{aligned}
    f^\infty&  = \frac{\pi^2}{kk'}
    \frac{e^{2\pi\eta}}{\left(e^{2\pi\eta}-1\right)\left(e^{2\pi\eta'}-1\right)} \chi_0 \eta\eta'  \\
    &\quad \times
    \Re \Big[
      {}_2F_1(i\eta,i\eta';1;\chi_0) \\
      &\qquad \times
      {}_2F_1(1-i\eta,1-i\eta';2;\chi_0)
    \Big].
  \end{aligned}
  \label{eqn:CB_pwsum_inf}
\end{equation}
Finite values of $f^{l_\text{max}}$ for higher-order terms of the interaction potential can still be calculated analytically; see the work of \citet{alder1956study} and \citet{chu1975rotational} for more detail.

Care must be taken when evaluating the hypergeometric functions numerically in (\ref{eqn:CB_dip_M}, \ref{eqn:CB_pwsum_inf}).
For example, $\chi_0$ is always strictly less than 0 for energetically allowed (de-)excitation and generally satisfies $\chi_0\ll-1$ as the scattering energy increases.
Linear transformations given by, e.g., \citet{pearson2017numerical}, allow one to directly evaluate the Taylor-series representation by ensuring $\abs{z}<1$, with the exception of several special cases, such as very near (de-)excitation which produce large values of $\eta$ or $\eta'$, leading to very poor convergence and numerical instability of the hypergeometric function Taylor series.

\subsection{Combining Cross Sections}

At this point, the cross sections have been obtained separately using the MQDT/R-matrix approach (Section \ref{sec:abinitio}) and the Coulomb-Born approximation, but have not been combined.
The electronic ground state of all three ions is a doublet state, so the R-matrix calculations are performed separately for the singlet and triplet spin multiplicities of the electron-ion system.
Cross sections are obtained for each spin multiplicity via (\ref{eqn:xs_mqdt}) and then averaged over the spin multiplicities, i.e.
\begin{equation}
  \sigma^{\text{Rmat}}_{i \to f}(E_\text{el}) =
  \frac{1}{4}
  \sigma^{(1)}_{i \to f}(E_\text{el})
  + \frac{3}{4} \sigma^{(3)}_{i \to f}(E_\text{el}),
  \label{eqn:xs_rmat_spinavg}
\end{equation}
where $i$ and $f$ are shorthand notation for the initial and final states, and the singlet/triplet cross sections are indicated by the superscripts $(1)$ and  $(3)$.
Both singlet and triplet calculations use the same truncated partial-wave basis of $l=0\ldotstwo5$.

The Coulomb-Born approach as described here does not resolve electron spin, so only two cross sections are calculated in the Coulomb-Born approximation: partial and total Coulomb-Born cross sections, calculated using (\ref{eqn:CB_xs_EinstA}) for $l_\text{max}$ taken as 5 and $\infty$, respectively.
The partial cross sections are calculated to mirror the partial-wave basis of the R-matrix calculations and ensure that no double-counting of low-$l$ partial waves occurs.
The three cross sections are combined to produce the so-called Coulomb-Born-corrected cross section,
\begin{equation}
  \sigma^\text{Tot}_{i \to f}(E_\text{el}) =
  \sigma^\text{Rmat}_{i\to f}(E_\text{el})
  + \sigma^\text{TCB}_{i\to f}(E_\text{el})
  - \sigma^\text{PCB}_{i\to f}(E_\text{el}).
  \label{eqn:CB_combine}
\end{equation}
The cross sections $\sigma^\text{PCB}_{i\to f}(E_\text{el})$ and $\sigma^\text{TCB}_{i\to f}(E_\text{el})$ correspond, respectively, to the partial and total Coulomb-Born cross sections.
The lower partial waves with $l=0\ldotstwo l_\text{max}$ are taken into account by the method described in Section \ref{sec:abinitio}, which better describes short-range effects than the Coulomb-Born approximation.
The contribution of the larger partial waves with $l>l_\text{max}$ are added in by considering the term $\sigma^\text{TCB}_{i\to f}(E_\text{el}) - \sigma^\text{PCB}_{i\to f}(E_\text{el})$, which effectively removes the $l=0\ldotstwo5$ contribution from the Coulomb-Born cross section.

\section{Results}
\label{sec:results}

\begin{figure}
  \centering
  \includegraphics[width=\linewidth]{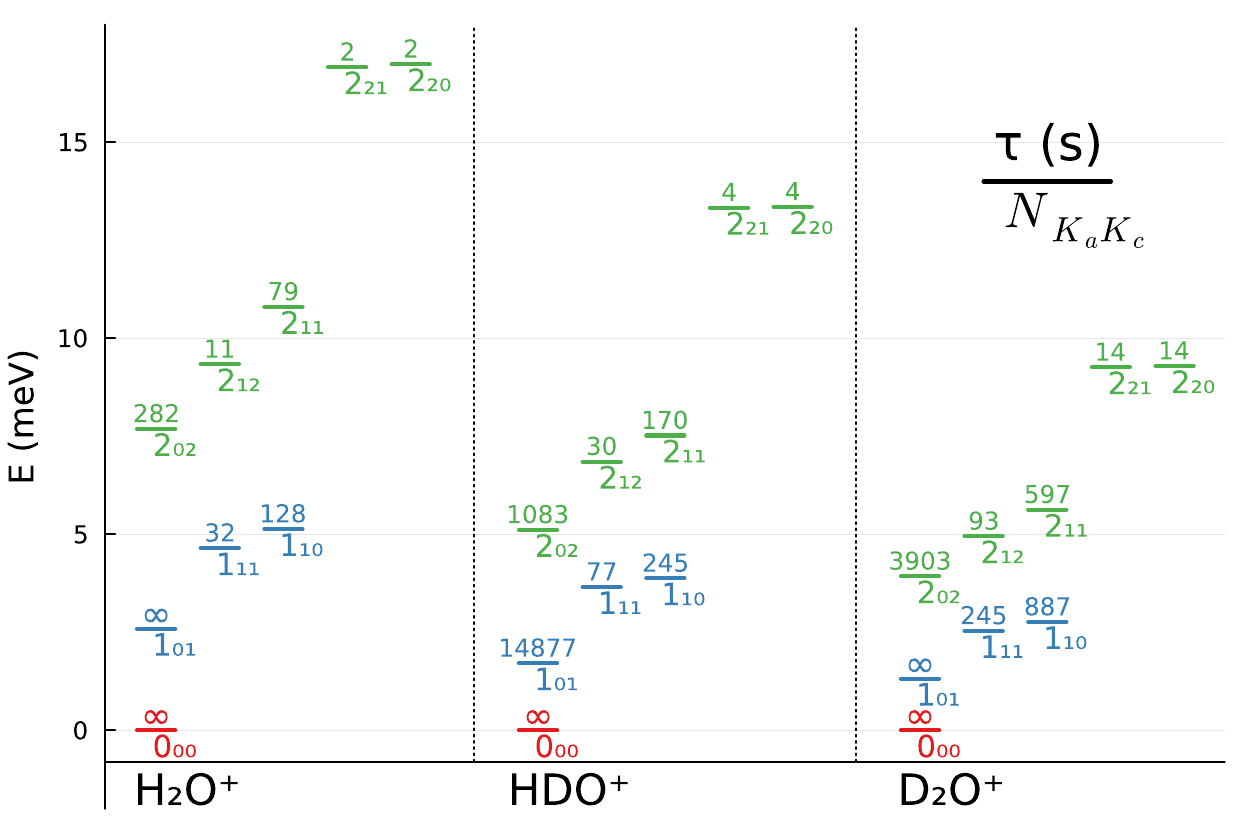}
  \caption{
    The lowest few rotational states of the \ce{H2O+}, \ce{HDO+}, and \ce{D2O+} molecular ions within the ground vibronic state.
    The states are represented as small horizontal bars plotted at their rotational energy, their radiative transition lifetime $\tau$ is given to the nearest integer in seconds above the bar, and below the bar is the rotational state label $N_{K_aK_c}$.
    Lifetimes are determined from the Einstein $A$ coefficient for radiative decay, which is zero for dipole-forbidden transitions.
    States that have no dipole-allowed decay channels are assigned the lifetime $\infty$.
    Horizontal spacing is added for legibility, and colors are used to further distinguish between states with different $N$.
  }
  \label{fig:states}
\end{figure}

\begin{figure*}
  \centering
  \begin{minipage}{0.48\linewidth}
    \includegraphics[width=\linewidth]{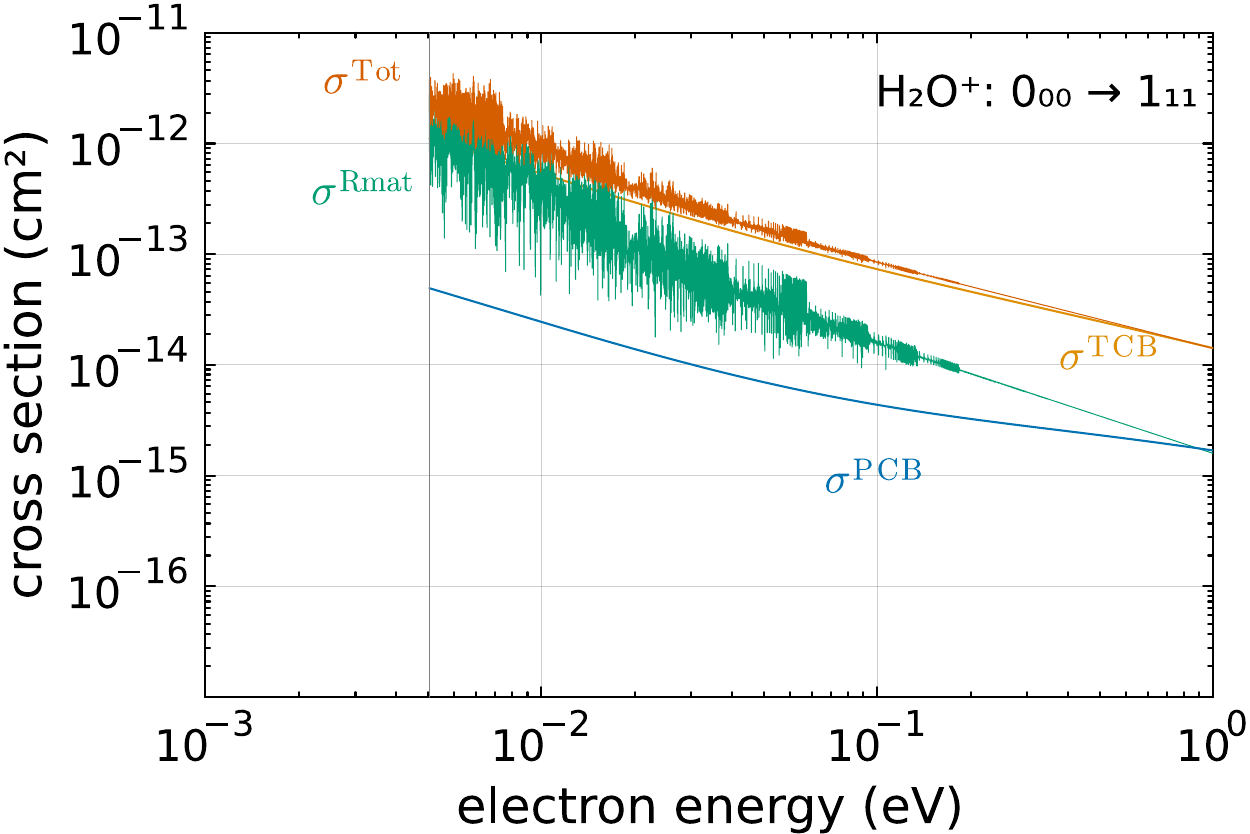}
  \end{minipage}
  \begin{minipage}{0.48\linewidth}
    \includegraphics[width=\linewidth]{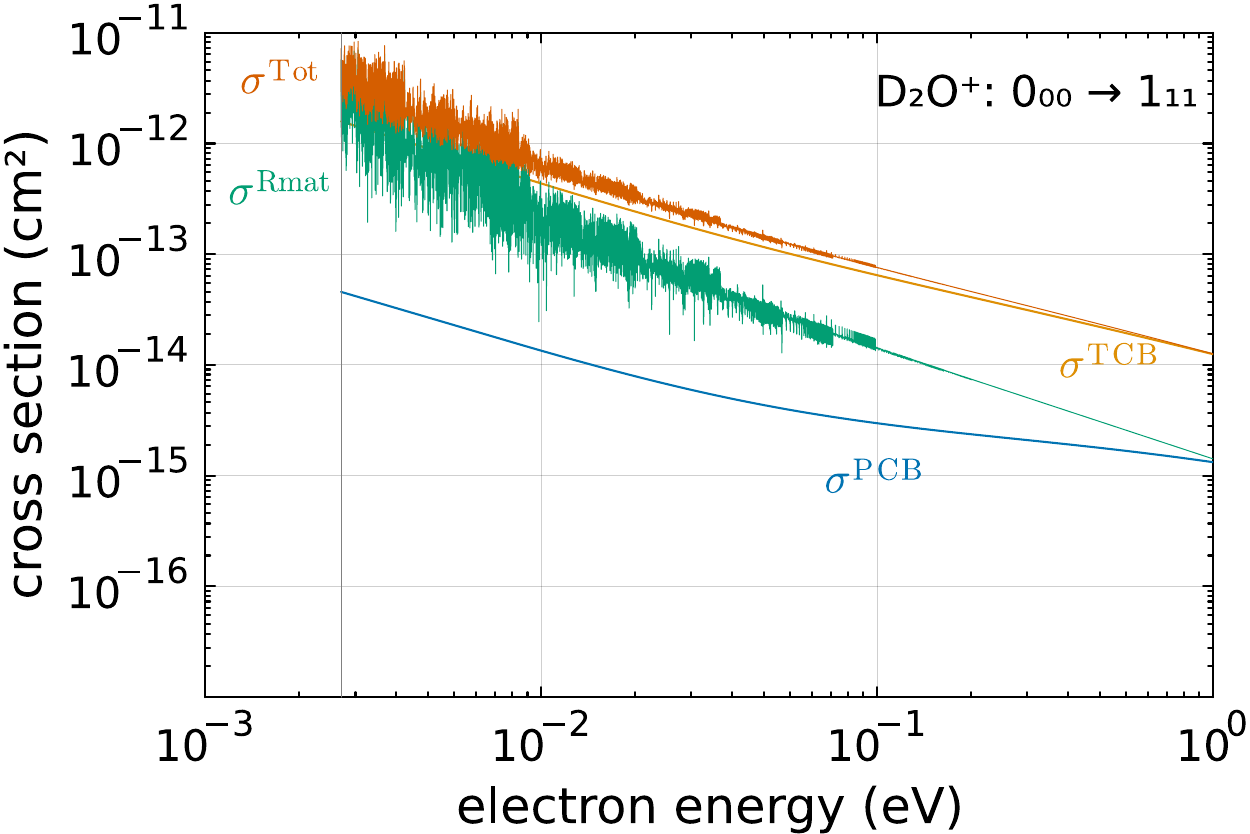}
  \end{minipage}
  \\
  \begin{minipage}{0.48\linewidth}
    \includegraphics[width=\linewidth]{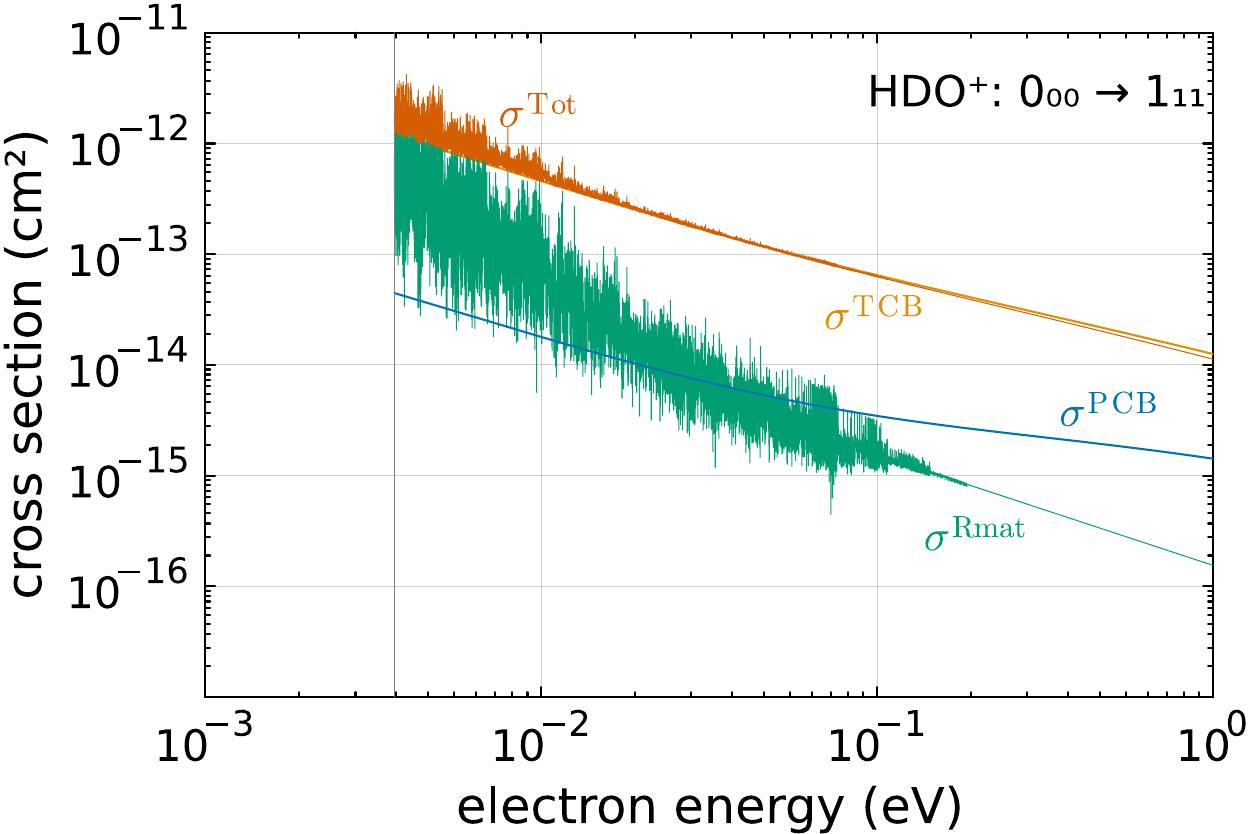}
  \end{minipage}
  \begin{minipage}{0.48\linewidth}
    \includegraphics[width=\linewidth]{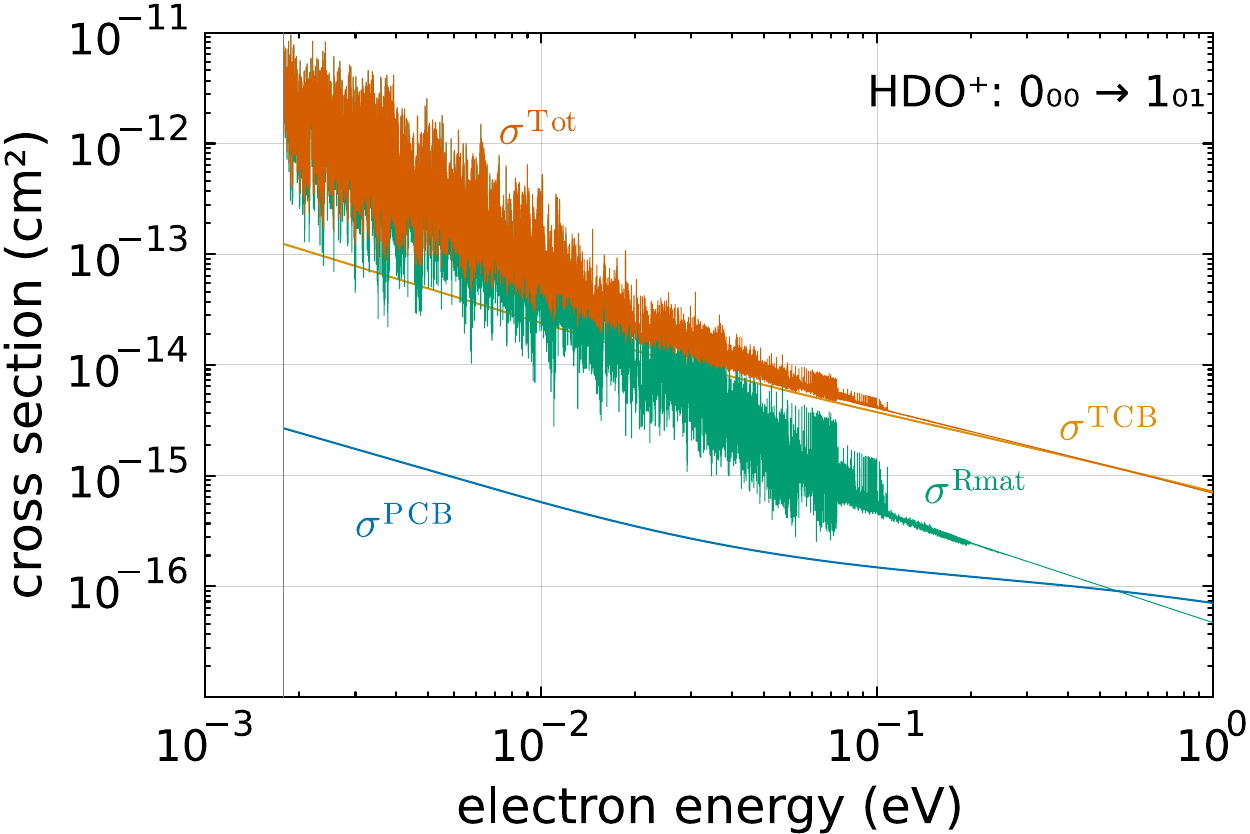}
  \end{minipage}
  \caption{
    The different cross sections for all dipole-allowed transitions in \ce{H2O+} (top left), \ce{D2O+} (top right), and \ce{HDO+} (bottom row) from the ground rotational state $0_{00}$, with each of the four cross sections in (\ref{eqn:CB_combine}) plotted.
    Note that \ce{HDO+} has two dipole-allowed transitions: $0_{00}\to1_{11}$ and $0_{00}\to1_{01}$.
    The rotational excitation threshold is denoted by a thin grey vertical line.
  }
  \label{fig:xs_combine}
\end{figure*}

\begin{figure}
  \centering
  \includegraphics[width=\linewidth]{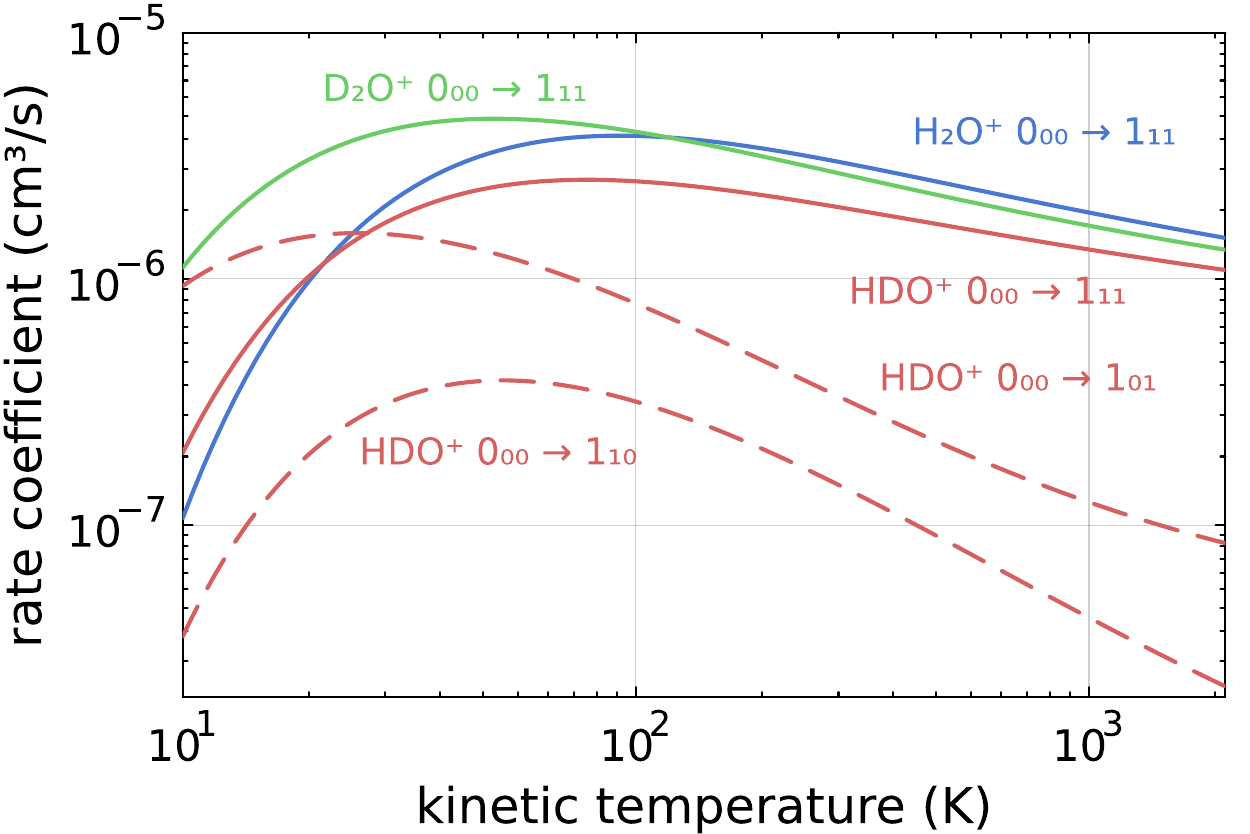}
  \caption{
    Rate coefficients for all allowed transitions from $N=0\to1$ for \ce{H2O+}, \ce{HDO+}, and \ce{D2O+}.
    Dashed lines represent transitions that are only allowed for \ce{HDO+} because of its lack of ortho/para symmetry.
  }
  \label{fig:rates_01}
\end{figure}

\begin{figure*}
  \centering
  \begin{minipage}{0.31\linewidth}
    \includegraphics[width=\linewidth]{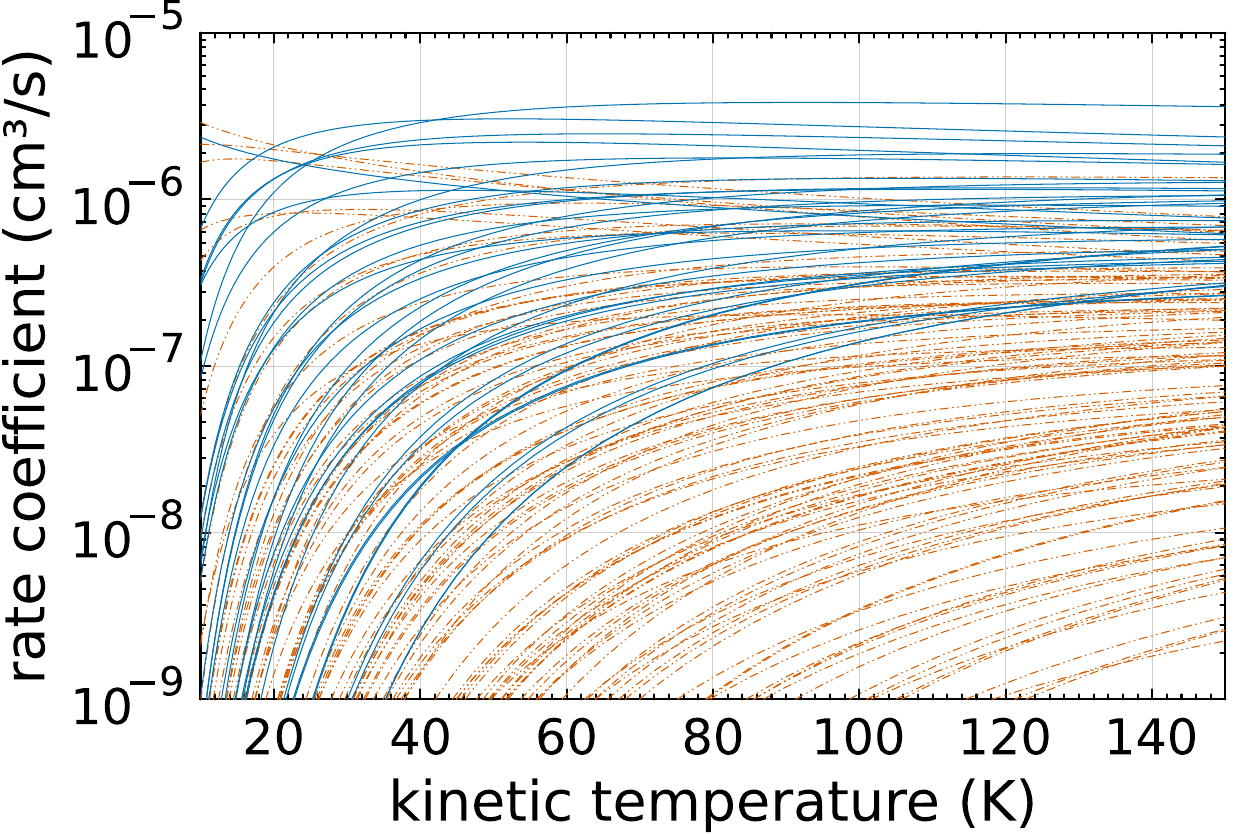}
  \end{minipage}
  \begin{minipage}{0.31\linewidth}
    \includegraphics[width=\linewidth]{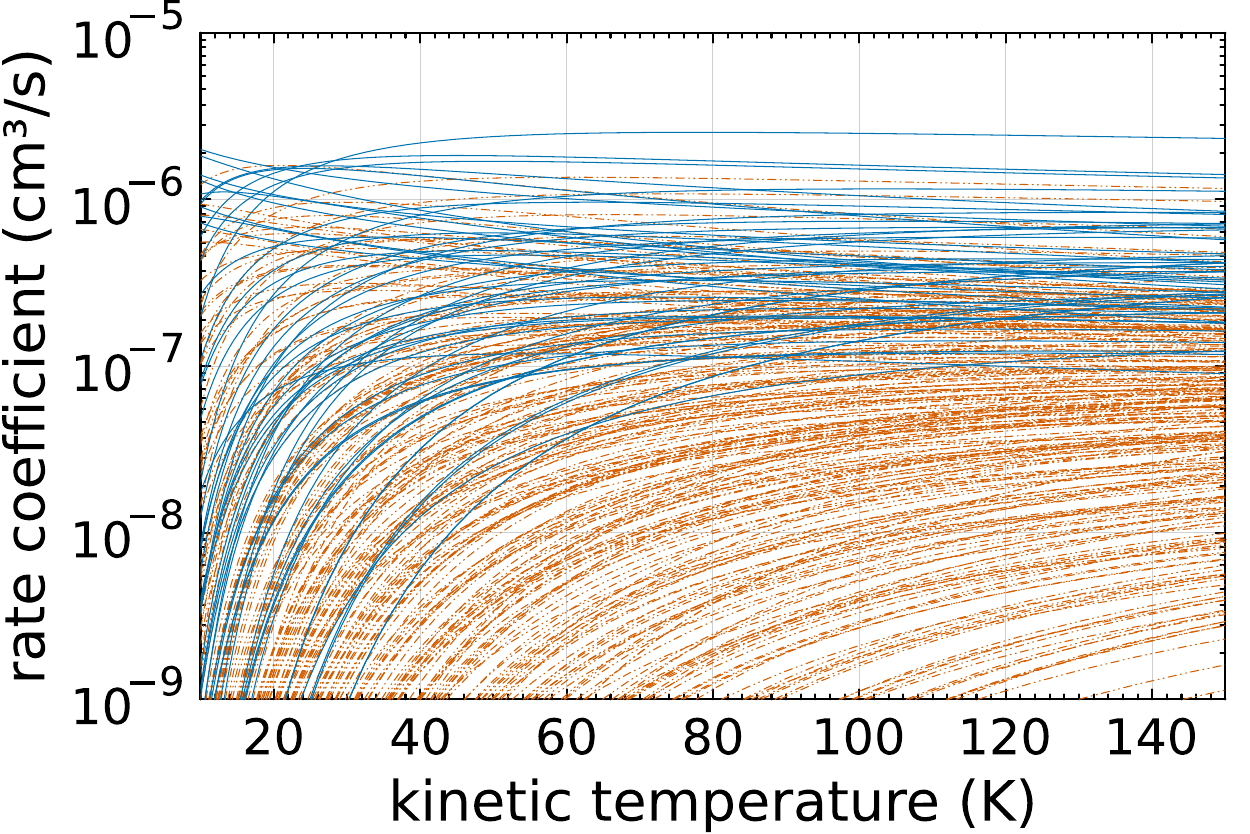}
  \end{minipage}
  \begin{minipage}{0.31\linewidth}
    \includegraphics[width=\linewidth]{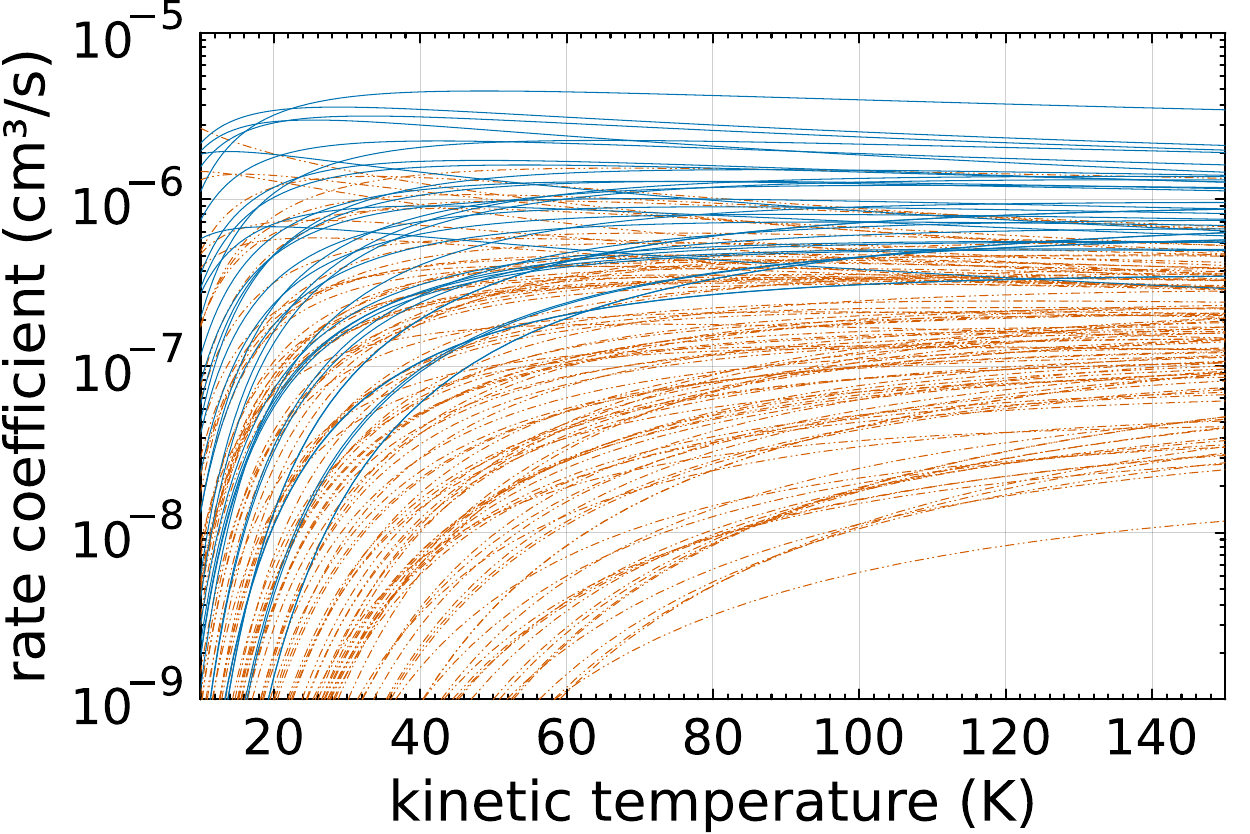}
  \end{minipage}
  \\
  \begin{minipage}{0.31\linewidth}
    \includegraphics[width=\linewidth]{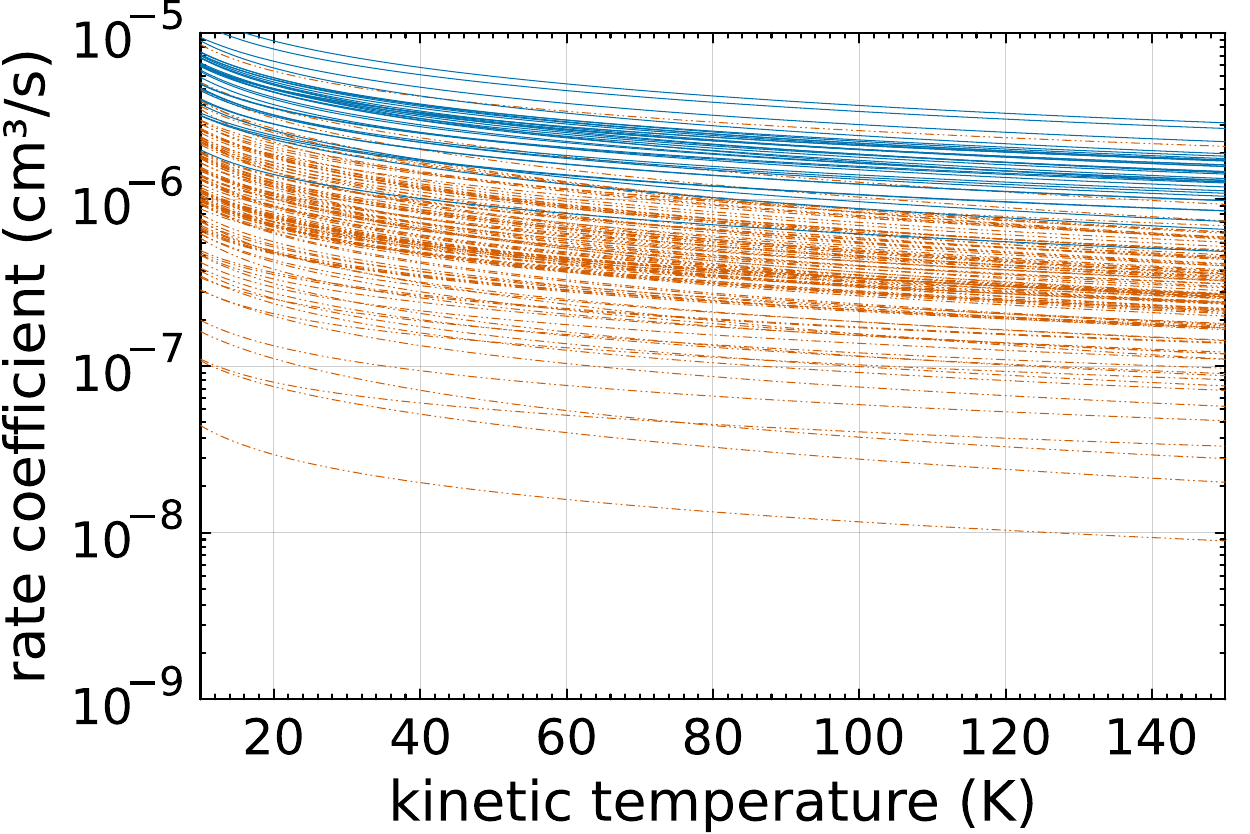}
  \end{minipage}
  \begin{minipage}{0.31\linewidth}
    \includegraphics[width=\linewidth]{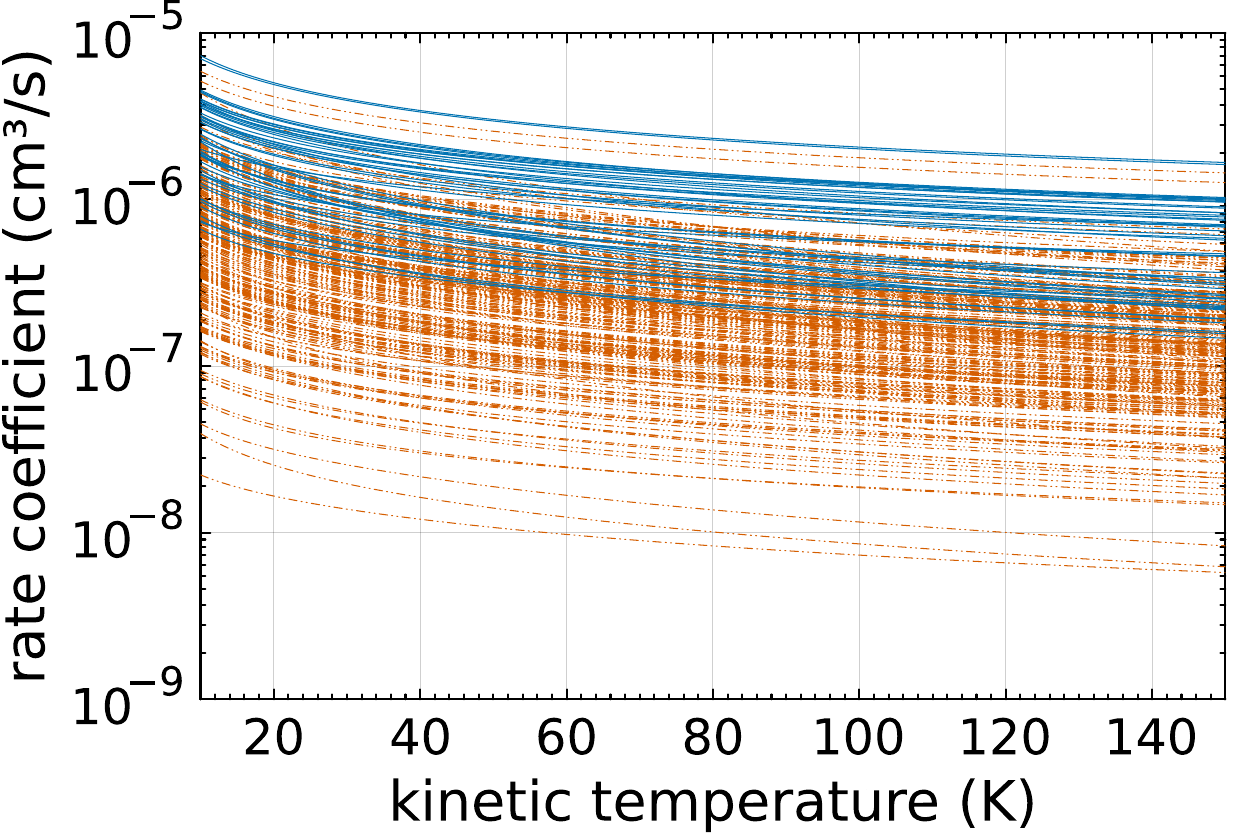}
  \end{minipage}
  \begin{minipage}{0.31\linewidth}
    \includegraphics[width=\linewidth]{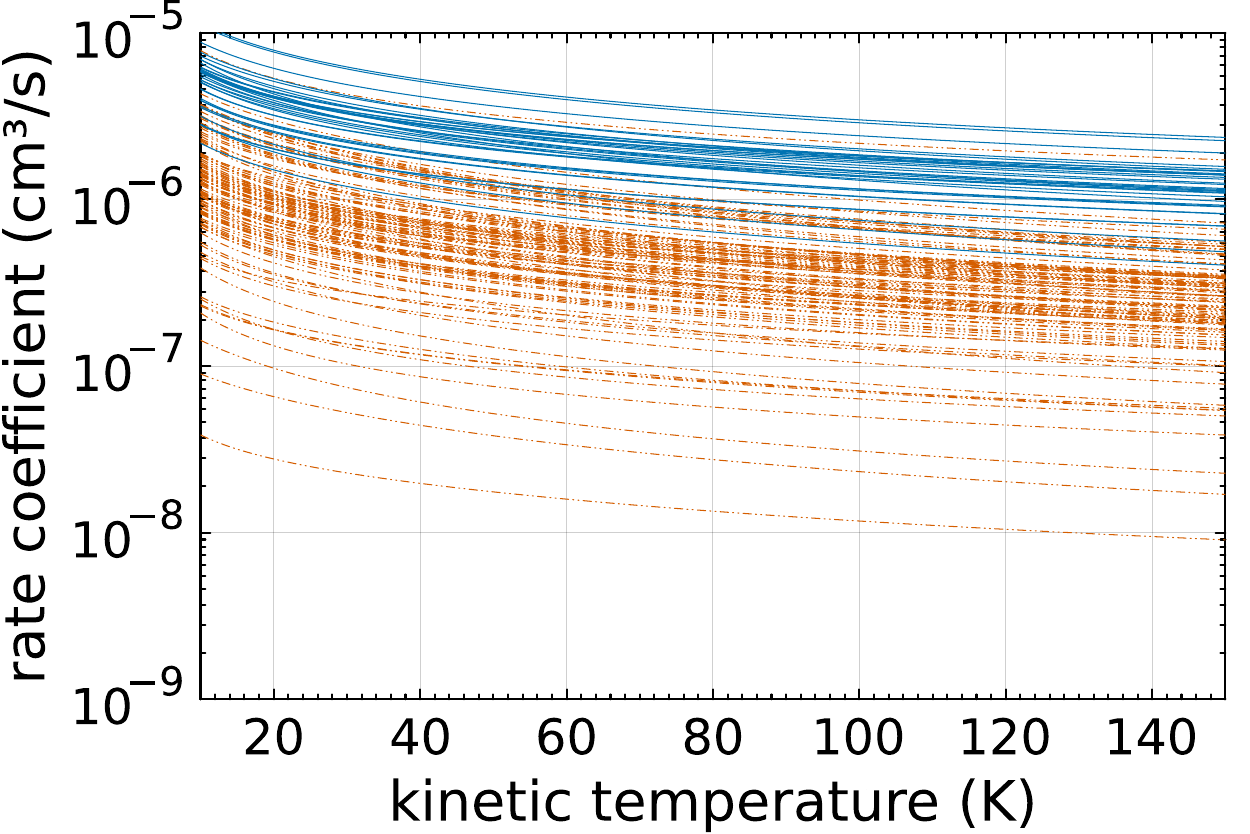}
  \end{minipage}
  \caption{
    Rotational excitation (upper row) and de-excitation (lower row) rate coefficients as a function of kinetic temperature for \ce{H2O+} (left), \ce{HDO+}, (middle), and \ce{D2O+} (right).
    Solid blue lines represent dipole-allowed transitions, dashed orange lines represent non-dipolar transitions.
  }
  \label{fig:all}
\end{figure*}

The rotational state energies follow the expected trend for \ce{H2O+}, \ce{HDO+}, and \ce{D2O+}.
As the molecules get heavier, the rotational constants (given in Table \ref{tab:target}) decrease, which results in energy levels that sit lower and closer together in energy, as seen in Figure \ref{fig:states}.
Additionally, for transitions that are allowed in all three molecules, their average radiative lifetimes increase with increasing mass because the Einstein $A$ coefficients are proportional to $\omega^3$ --- the cubed angular frequency of the emitted photon.
As the energy levels get closer, $\omega$ decreases and the average radiative lifetimes $\tau = 1/A$ increase.

Figure \ref{fig:xs_combine} demonstrates all four cross sections in (\ref{eqn:CB_combine}) for the dipolar $N=0\to1$ transitions in all three ions, and generally reproduces the behavior observed in a similar study applied to \ce{CH+}\cite{forer2024kinetic}, a light diatomic ion with a relatively strong electric dipole moment ($\sim$1.7~Debye).
At lower energies, the \quotes{Rmat} (no Coulomb-Born correction) cross sections feature many resonances.
These are Rydberg resonances associated with the closed rotational channels of the ion, which become less and less numerous as channels open when the electron energy increases.
At the lowest energies, the \quotes{Rmat} cross sections provide a noticeable enhancement to the \quotes{TCB} cross sections because of the effect of the Rydberg resonances, especially for the $0_{00}\to1_{11}$ transition in \ce{HDO+} because its dipole is mostly aligned with the $B$-axis (see Table \ref{tab:target} for electronic dipole moment values).
At higher energies, the \quotes{Rmat} cross section's resonances die out; most rotational channels are open and, given that this approach ignores the vibrational motion of the nuclei and energy dependence of the body-frame K-matrices, the cross sections become approximately straight lines in logarithmic space (up to very small resonances and numerical artifacts) which behave as $1/E_\text{el}$.
In this regime, the Coulomb-Born cross sections clearly dominate, which is to be expected given the nature of the problem.
Born-type cross sections are generally understood as a high-energy approximation --- that the deviation of the interaction potential between the target ion and the electron is only a perturbation from the Coulomb potential.

Rate coefficients can be obtained by assuming a thermalized electron population and averaging the cross section over a Maxwell-Boltzmann distribution in electron speed \cite{forer2024kinetic}.
Such state-resolved coefficients (i.e., not averaged over initial or final rotational states) for the $N=0\to1$ transition are shown for all three ions in Figure (\ref{fig:rates_01}).
The \ce{HDO+} ion lacks the nuclear permutation symmetry engendered by the identical nuclei of \ce{H2O+} and \ce{D2O+}, which is why it has three $N=0\to1$ transitions.
The additional transitions that are seen in \ce{HDO+} would be between two states of different parity for \ce{H2O+} and \ce{D2O+}, and are forbidden.
These extra transitions in \ce{HDO+} reduce the overall flux available to the $0_{00}\to1_{11}$ transition, which is likely why this transition has the smallest magnitude in the case of \ce{HDO+}.

At higher temperatures, the $0_{00}\to1_{11}$ transition is the strongest $N=0\to1$ transition in \ce{HDO+} because this transition corresponds to a $B$-type dipolar transition (see Table \ref{tab:selection} for dipole selection rules) and the dipole moment along the $B$ axis is by far the most significant, as seen in Table \ref{tab:target}.
The next largest $N=0\to1$ transition in \ce{HDO+} is $0_{00}\to1_{01}$, which is also a dipole-allowed transition for this molecule because it has a nonzero dipole component along the $A$ axis.
Finally, the non-dipolar transition $0_{00}\to1_{10}$ is the smallest.
At lower temperatures, the relative magnitude of the rate coefficients for all transitions changes significantly due to the shifted thresholds for rotational excitation.
Below 10~K (not plotted), the $0_{00}\to1_{01}$ transition is the largest of all $N=0\to1$ transitions available for the three ions because its excitation threshold is the first to open as the collision energy increases, as seen in Figure \ref{fig:states}.

Figure \ref{fig:all} plots rate coefficients for all calculated transitions for all three ions for rotational excitation (top row) and de-excitation (bottom row).
The solid blue lines represent dipole-allowed rate coefficients (and are therefore enhanced by the dipolar Coulomb-Born approximation), while the dashed orange lines represent cross sections that are dipole-forbidden and are therefore obtained entirely from the rotationally resolved S-matrix (\ref{eqn:xs_mqdt},\ref{eqn:xs_rmat_spinavg}).
The overall trend for these ions is that the dipole-allowed rate coefficients to dominate.
There are several dipole-forbidden transitions that have significant magnitude, coming typically from de-excitation cross sections between states that are very close in energy.
Such states tend to have large rate coefficients at lower kinetic temperatures, as noticeable in Figure \ref{fig:rates_01}.
Figure \ref{fig:all} also demonstrates the larger number of transitions in \ce{HDO+} given the lack of ortho/para separation.

\section{Conclusion}
\label{sec:conclusion}

A theoretical development for electron-impact rotational excitation of asymmetric top molecular ions that combines R-matrix theory, MQDT, frame transformation theory, and the Coulomb-Born approximation is presented in this paper and applied to the three isotopologues \ce{H2O+}, \ce{HDO+}, and \ce{D2O+} for all rotational transitions between the $N=0$ and $N=4$ states.
While this work develops the Coulomb-Born approximation for all orders of the multipole expansion of the interaction potential between the target ion and the scattering electron, only the dipole term is used, which is expected to provide the dominant contribution as in the case of several other ions\cite{faure2001electron} with strong electric dipole moments.
Dipole-allowed transitions are found to dominate in general for all three species, but dipole-forbidden transitions show significant magnitude in state-selected kinetic temperature rate coefficients, especially at lower kinetic temperatures.
In general, the Coulomb-Born approximation is responsible for the largest contribution in rotational (de-)excitation cross sections and rate coefficients when the transitions are dipole-allowed (and when the electric dipole component has significant magnitude).
This is especially true at higher kinetic temperatures, suggesting that the Coulomb-Born approximation can be reliably used while neglecting short-range effects --- as is expected of a Born-type formulation.
However, especially at lower kinetic temperatures, the R-matrix calculations can provide a significant enhancement in the rotational (de-)excitation cross sections and rate coefficients due to the effects of densely packed rotational Rydberg resonances, and also allow for the calculation of transitions that are dipole-forbidden.
Many of these dipole-forbidden transitions show appreciable magnitude when compared to the dipole-allowed transitions, but at higher energies and kinetic temperatures, the Coulomb-Born approximation is likely sufficient for these ions due to their strong electric dipole moment.

This approach can be extended to symmetric-top molecular ions, including non-polar ions (which may still have a significant quadrupole moment and short-range scattering effects), to vibrating targets, and to neutral targets.
In the case of neutrals, the Born closure (also known as Born completion) takes on a slightly different form owing to the different asymptotic radial functions in the electronic coordinate, as discussed by \citet{feldt2008analytic}, while the angular components coming from the rotational motion of the target molecule remain the same.
Additionally, the frame transformation can be extended to include the energy dependence of the body-frame matrices, as discussed by \citet{gao1989energy,hvizdos2025competing}.

\begin{acknowledgments}
  This work was funded by NSF award no. AST-2303895.
  Many thanks to Alexandre Faure at Université Grenoble Alpes, Viatcheslav Kokoouline at the University of Central Florida, and Daniel W. Savin at Columbia University for helpful discussions, as well as to Dávid Hvizdoš for providing the computer code for the R-matrix elimination and Whittaker function evaluation.
\end{acknowledgments}

\section*{Data Availability Statement}
All state-selected kinetic temperature rate coefficients will be provided as supplementary material to this work and will be made available on the Excitation of Molecules and Atoms for Astrophysics (EMAA) database \cite{faure2025excitation}.

\appendix
\setcounter{table}{0}
\renewcommand{\thetable}{\Alph{section}\arabic{table}}

\section{The Rotational Hamiltonian}
\label{app:rot}

The Watson asymmetric-top (A-reduced) effective rotational Hamiltonian \cite{watson1967determination} is given to fourth order by
\begin{align}
  H_\text{rot} &= B_x\hat{N}_x^2 + B_y\hat{N}_y^2 + B_z\hat{N}_z^2
  \label{eqn:app:HRR}
  \\
  &\begin{aligned}
    & -\Delta_N\hat{\boldsymbol N}^4
    - \Delta_{NK}\hat{\boldsymbol N}^2\hat{N}^2_z
    - \Delta_K\hat{N}^4_z
    \\
    & \quad
    - \frac{1}{2} \left[
      \delta_N\hat{\boldsymbol N}^2 + \delta_K\hat{N}^2_z,
      \hat{N}^2_+ + \hat{N}^2_-
    \right]_+ + \boldsymbol{O}(\hat{\boldsymbol N}^6)
  \end{aligned}
  \label{eqn:app:HCD4}
\end{align}
where $\hat{N}_\pm = \hat{N}_x \pm i \hat{N}_y$,
$B_{\{x,y,z\}}$ are the rotational constants with respect to the orthogonal molecule-fixed axes $\{\hat{x},\hat{y},\hat{z}\}$, and the remaining coefficients are to be determined by some combination of experimental, computational, or observational efforts.
It is an effective Hamiltonian that includes the rigid-rotor Hamiltonian (\ref{eqn:app:HRR}) and a correction term (\ref{eqn:app:HCD4}) that accounts for centrifugal distortion effects arising from the molecule's rotational motion.
In principle, higher-order corrections (e.g., sextic, octic..) exist, but are typically only used for high-accuracy spectroscopy.
The centrifugal distortion parameters depend on the choice in axes, and the convention for the A-reduction is that the $\hat{z}$ and $A$ axes are taken to be aligned.
The fourth order centrifugal distortion coefficients used here can be found for each molecule in Table \ref{tab:target}.
However, given that the $C_{2v}$ scattering calculations in \texttt{UKRmol+} are restricted to have the $C_2$ symmetry axis for \ce{H2O+} and \ce{D2O+} along the $\hat{z}$ axis, this results in two separate reference frames.
Therefore, the rotational wavefunctions for \ce{H2O+} and \ce{D2O+} (\ref{eqn:wf_rot}) obtained by diagonalizing the above $H_\text{rot}$ are rotated to be in the same reference frame as the electron scattering calculations via the $zyz$-rotation
\begin{equation}
  V^b = D^N\left(0, \frac{\pi}{2} , \frac{\pi}{2} \right)^\dagger V^a,
  \label{eqn:app:rotate_eigvecs}
\end{equation}
where $V^a$ is the square matrix whose columns are the $2N+1$ eigenvectors of $H_\text{rot}$ for a given value of $N$, $D^{N\dagger}(\alpha,\beta,\gamma)$ is Hermitian adjoint of the Wigner $D$-matrix with Euler angles $(0,\pi/2,\pi/2)$, and $V^b$ is the desired matrix of rotated eigenvectors with the $z$-axis coinciding with the $C_2$ axis.

\section{Details of the Coulomb-Born Approximation}
\label{app:CB}

The molecule-fixed frame (MF) perturbation to the Coulomb potential is given by $H_\text{MF}'$ (\ref{eqn:potv_multipoles}--\ref{eqn:multipoles}).
The MF angular component can be expressed in terms of the rotated space-fixed frame (SF) angular component, i.e.
\begin{equation}
  Y_\xi^\mu(\chi,\psi) =
  \sum\limits_{\nu=-\xi}^\xi D^\xi_{-\nu,\mu}(\alpha,\beta,\gamma)Y_\xi^\nu(\theta,\phi),
  \label{eqn:app:CB_YLMrot}
\end{equation}
where $(\theta, \phi)$ are SF angles, $(\alpha,\beta,\gamma)$ are Euler angles, and $D^{\xi}_{-\nu,-\mu}(\alpha,\beta,\gamma)$ is an element of the Wigner D-matrix.
The perturbation $H_\text{MF}'$ can be redefined in terms of SF angles as $H_\text{SF}'$,
\begin{equation}
  \begin{aligned}
    H'_\text{SF} &= \sum\limits_{\xi=1}^\infty
    \sqrt{\frac{4\pi}{2\xi+1}}
    \sum\limits_{\mu=-\xi}^\xi
    \frac{Q_{\xi\mu}}{r^{\xi+1}}  \\
    &\quad \times \sum\limits_{\nu=-\xi}^\xi
    D^\xi_{-\nu,-\mu}(\alpha,\beta,\gamma) Y_\xi^\nu(\theta,\phi).
  \end{aligned}
  \label{eqn:app:CB_HSF}
\end{equation}
This work only aims to produce integrated cross sections, which are obtained by
\begin{equation}
  \begin{aligned}
    \sigma_{N\tau\to N'\tau'} &=
    \frac{1}{4\pi^2} \frac{k'}{k} \int\limits_{\hatp{k}} d\hatp{k}
    \sum\limits_{MM'} \frac{1}{2N+1} \\
    &\qquad \times \abs{\braket{N'\tau'M'k'|H'_\text{SF}|N\tau Mk}}^2,
  \end{aligned}
  \label{eqn:app:CB_xs_unsolved}
\end{equation}
where $\ket{N\tau Mk}$ are the asymptotic wavefunctions that are taken to be separable into asymmetric-top wavefunctions $\ket{N\tau M}$ (\ref{eqn:wf_rot}) and Coulomb wavefunctions $\ket{k}$,
\begin{equation}
  \begin{gathered}
    \ket{k} = \sum\limits_{lm} 4\pi (-1)^m i^l e^{i\sigma_l} Y_l^{-m}(\hat{k}) Y_l^{m}(\hat{r}) (kr)^{-1} F_l(kr),
    \\
    \ket{k'} = \sum\limits_{l'm'} 4\pi (-1)^{m'} i^{l'} e^{-i\sigma_{l'}} Y_{l'}^{-m'}(\hatp{k}) Y_{l'}^{m'}(\hat{r}) (k'r)^{-1} F_{l'}(k'r),
    \\
    \sigma_l = \arg\left[\Gamma(l + 1 + i\eta)\right], \quad \eta = -Z/k,
  \end{gathered}
  \label{eqn:k}
\end{equation}
where initial and final states are denoted, respectively, by unprimed and primed quantities, $F_l(kr)$ is the energy-normalized radial Coulomb function, $Z$ is the target's total electric charge, and $\vec{k}$ is the electron wavevector.

The quantity $\braket{N'\tau'M'k'|H_\text{SF}'|N\tau Mk}$ can be factored as
\begin{equation}
  \begin{aligned}
    A &\equiv \braket{N'\tau'M'k'|H_\text{SF}'|N\tau Mk}
    \\ &
    =\braket{N'\tau'M'|\braket{k'|H_\text{SF}'|k}|N\tau M}
    \\ &
    =\braket{{N'\tau'M'|I(\alpha,\beta,\gamma)|N\tau M}},
  \end{aligned}
  \label{eqn:app:H_SF_avg}
\end{equation}
where
\begin{equation}
  \begin{aligned}
    &I(\alpha,\beta,\gamma) \equiv \braket{k'|H_\text{SF}'|k}
    \\
    &= \frac{(4\pi)^{5/2}}{\sqrt{2\xi+1}}
    \sum\limits_{\xi}
    \sum\limits_{\mu\nu}
    \sum\limits_{ll'}
    \sum\limits_{m'}
    (-1)^{m'} i^{l - l'} e^{i(\sigma_l + \sigma_{l'})}
    Q_{\xi\mu}
    \\
    & \quad \times
    D^\xi_{-\nu,-\mu}(\alpha,\beta,\gamma)
    \sqrt{\frac{2l+1}{4\pi}}
    Y_{l'}^{-m'}(\hat{k'})
    \\
    & \quad \times
    \int d\hat{r} Y_\xi^\nu(\hat{r}) Y_l^0(\hat{r}) Y_{l'}^{m'}(\hat{r})^*
    \\
    & \quad \times
    \int r^2 dr \frac{1}{kk'r^2} \frac{1}{r^{\xi + 1}} F_l(kr) F_{l'}(k'r) ,
  \end{aligned}
  \label{eqn:app:I_explicit}
\end{equation}
and the last two terms are the angular and radial integrals.
The sum over $m$ collapses to only the $m=0$ term when the incident electron's wavevector is taken to be aligned with the SF $\hat{z}$-axis.
The integral over three spherical harmonics can be expressed in terms of the Wigner 3-j symbols, which reduces (\ref{eqn:app:I_explicit}) to
\begin{equation}
  \begin{aligned}
    &I(\alpha,\beta,\gamma) =
    \frac{(4\pi)^{5/2}}{2\xi+1}
    \sum\limits_{\xi}
    \sum\limits_{\mu\nu}
    Q_{\xi \mu}
    D^\xi_{-\nu,-\mu}(\alpha, \beta, \gamma) \\
    &\times
    \sum\limits_{ll'}
    \sum\limits_{m'}
    i^{l - l'} e^{i(\sigma_l + \sigma_{l'})}
    \wignerjjj{l}{l'}{\xi}{0}{0}{0} \wignerjjj{l}{l'}{\xi}{0}{-m'}{\nu}
    \\
    &\times
    (2l+1)\frac{\sqrt{(2\xi+1)(2l'+1)}}{4\pi}
    M^\xi_{ll'}
    Y_{l'}^{-m'}(\hat{k'}).
  \end{aligned}
  \label{eqn:app:I_reduced}
\end{equation}

Using the orthogonality relations of three Wigner D-matrices in terms of the 3-j symbols when integrating over all Euler angles, one can then re-express the matrix element (\ref{eqn:app:H_SF_avg}) as
\begin{equation}
  \begin{aligned}
    &A = \sqrt{(2N+1)(2N'+1)}(4\pi)^{3/2}
    \\ &\times
    \sum\limits_{\xi}
    \sum\limits_{\mu\nu}
    \sum\limits_{KK'}
    (-1)^{M}
    Q_{\xi\mu}
    G^{\xi\mu}_{N\tau,N'\tau'},
    \wignerjjj{N}{N'}{\xi}{-M}{M'}{-\nu}
    \\ &\times
    \sum\limits_{ll'm'}
    i^{l - l'} e^{i(\sigma_l + \sigma_{l'})}
    \wignerjjj{l}{l'}{\xi}{0}{0}{0} \wignerjjj{l}{l'}{\xi}{0}{-m'}{\nu}
    M^\xi_{ll'}
    \\ &\times
    Y_{l'}^{-m'}(\hat{k'}),
  \end{aligned}
\end{equation}
where $G^{\xi\mu}_{N\tau,N'\tau'}$ is defined as in (\ref{eqn:CB_G}), i.e.
\begin{equation}
  G^{\xi\mu}_{N\tau,N'\tau'} =
  \sum\limits_{KK'}
  (-1)^K
  c^{(N\tau)}_K c^{(N'\tau')*}_{K'}
  \wignerjjj{N}{N'}{\xi}{-K}{K'}{-\mu}.
  \label{eqn:app:CB_G}
\end{equation}
The quantity $A$ is not to be confused with the Einstein $A$ coefficient; the total integrated cross section (\ref{eqn:app:CB_xs_unsolved}) depends on $\abs{A}^2=A^*A$ and the sum over the SF projections $M,M'$.
Together, these introduce several orthogonality relations that simplify the integrated cross section,
\begin{equation}
  \begin{aligned}
    \sigma_{N\tau\to N'\tau'} &=
    \frac{1}{4\pi^2} \frac{k'}{k} \int\limits_{\hatp{k}} d\hatp{k}
    \sum\limits_{MM'} \frac{1}{2N+1} A^*A,
  \end{aligned}
  \label{eqn:app:CB_xs_unsolved_A}
\end{equation}
and allows one to obtain the final expression for the total integrated cross section in the Coulomb-Born approximation (\ref{eqn:CB_xs_solved}).

\section{Matrix Elements of the Rotational Frame Transformation}
\label{sec:app:rft}

In the space-fixed (SF) frame, one can define a roelectronic channel as
\begin{equation}
  \ket{\text{SF}} = \ket{\text{SF};NM_N\tau nlm} = \ket{\text{SF};NM_N\tau} \otimes \ket{nlm}.
  \label{eqn:app:SF_ket}
\end{equation}
Using the expansion of the asymmetric-top rotational eigenfunctions (\ref{eqn:wf_rot}) in the symmetric top basis, an SF channel can be re-expressed as
\begin{equation}
  \begin{aligned}
    \ket{\text{SF}} &= \sqrt{\frac{2N+1}{8\pi^2}}
    \sum\limits_{K\lambda} c^{(N\tau)}_K
    Y_l^\lambda(\theta,\phi)
    \\&
    \sum\limits_{J}
    \sum\limits_{M_J}
    \sum\limits_\Omega
    C^{JM_J}_{NM_N,lm}
    C^{J\Omega}_{NK,l\lambda}
    D^{J*}_{M_J,\Omega}(\alpha,\beta,\gamma).
  \end{aligned}
\end{equation}
which makes use of the same rotation mentioned in \ref{eqn:app:CB_YLMrot} and the product of two Wigner D-matrices,
\begin{equation}
  \begin{aligned}
    &D^{N*}_{M_N,K}(\alpha,\beta,\gamma) D^{l*}_{m,\lambda}(\alpha,\beta,\gamma)
    =
    \\ &
    \sum\limits_{J=\abs{N-l}}^{N+l}
    \sum\limits_{M_J=-J}^{J}
    \sum\limits_{\Omega=-J}^{J}
    C^{JM_J}_{NM_N,lm} C^{J\Omega}_{NK,l\lambda}
    D^{J*}_{M_J,\Omega}
  \end{aligned}
  \label{eqn:Dprod}.
\end{equation}
The quantities $M_N$, $M_J$, and $m$ are the projections of $\vec{N}$, $\vec{J}$, and $\vec{l}$ on the SF $\hat{z}$-axis, respectively.
The electronic S-matrix is expressed in a basis of electronic channels $\ket{nl\lambda}$, and
the goal is to arrive at SF rotationally resolved channels.
Therefore, the intermediate molecule-fixed (MF) roelectronic channels,
\begin{equation}
  \ket{\text{MF};JM_J\Omega nl\lambda} = \sqrt{\frac{2J+1}{8\pi^2}} D^{J*}_{M_J,\Omega}(\alpha,\beta,\gamma) \ket{nl\lambda},
  \label{eqn:app:MF_ket}
\end{equation}
are introduced, where the rotational part is built from Wigner $D$-matrices.
It should be stressed that the asymmetric-top character of the target is contained in the coefficients $c_K^{(N\tau)}$, and not in this choice of intermediate basis.
The overlap between the SF and MF channels for fixed $\{JM_J\Omega nl\lambda\}$ is given by
\begin{equation}
  \begin{aligned}
    &\braket{\text{MF}|\text{SF}} \equiv \braket{\text{MF};JM_J\Omega nl\lambda | \text{SF};NM_N\tau nlm} =\\
    &\quad =
    \sqrt{\frac{2N+1}{2J+1}}\,
    C^{J M_J}_{N M_N,\, l m}
    \sum_{K}
    c_K^{(N\tau)}
    C^{J\Omega}_{N K,\, l \lambda}.
  \end{aligned}
  \label{eqn:app:MF_SF}
\end{equation}
The above is useful when defining the rotationally resolved SF S-matrix via frame transformation,
\begin{equation}
  S^{\text{SF}}_{i f}
  =
  \sum_{\text{MF},\,\text{MF}'}
  \braket{\text{SF}_i \mid \text{MF}}\,
  S_{\text{MF},\text{MF}'}\,
  \braket{\text{MF}' \mid \text{SF}_f},
\end{equation}
where $ S_{\text{MF},\text{MF}'} = \braket{\text{MF}|\hat S^{\text{elec}}|\text{MF}'}$.
Initial and final states are indicated with the $i$ and $f$ subscripts.
Within the fixed-nuclei frame-transformation framework, $\hat{S}^{\text{elec}}$ acts only in the MF electronic channel basis and carries no dependence on the Euler angles or on the MF projection labels.
Consequently,
\begin{equation}
  \braket{\text{MF}|\hat{S}^{\text{elec}}|\text{MF}'}
  =
  \delta_{J J'}\,
  \delta_{M_J M_J'}\,
  \delta_{\Omega \Omega'}\,
  S_{n l \lambda,\, n' l' \lambda'}.
\end{equation}
By orthogonality of the Wigner $D$-matrices, only single summations over $J$, $M_J$, and $\Omega$
remain in the final transformation.

The integrated cross section in terms of the SF S-matrix is
\begin{equation}
  \sigma_{(N\tau\to N'\tau')}
  = \frac{\pi}{2m_\text{e}E_\text{el}}
  \frac{1}{2N+1}
  \sum\limits_{\substack{mm'\\M_NM_N'}}
  \abs{I - S^\text{SF}}^2,
  \label{eqn:app:ICS_SMAT1}
\end{equation}
which reduces to
\begin{equation}
  \sigma_{N\tau\to N'\tau'}
  =
  \frac{\pi}{2m_\text{e}E_\text{el}}
  \sum\limits_{J}
  \frac{2J+1}{2N+1}
  \sum\limits_{ll'}
  \abs{I - S^{J}}^2,
  \label{eqn:app:ICS_SMAT2}
\end{equation}
via Clebsch-Gordan orthogonality.
The matrix $S^J$ is subtracted from the identity matrix $I$ to remove the contribution of the incoming wave in the case of elastic scattering; it is included only at this point because the above steps do not change in arriving at the above, which is the same as (\ref{eqn:xs_mqdt}) when $S^J$ is taken to be the \textit{physical} S-matrix that is obtained via the channel elimination procedure (\ref{eqn:CCEP}),
\begin{equation}
  \hspace{-.5em}
  \begin{aligned}
    &S^{J}_{N\tau nl,N'\tau'n'l'} = \frac{\sqrt{(2N+1)(2N'+1)}}{2J+1}
    \sum\limits_{\Omega}
    \\& \times
    \sum\limits_{\lambda\lambda'}
    \sum\limits_{KK'} c^{(N\tau)*}_{K} c^{(N'\tau')}_{K'} C^{J\Omega}_{NK,l\lambda} C^{J\Omega}_{N'K',l'\lambda'}
    S_{nl\lambda,n'l'\lambda'},
  \end{aligned}
  \hfill
  \label{eqn:app:RFT}
\end{equation}
which is identical to (\ref{eqn:RFT}) and can also be expressed as
\begin{equation}
  \begin{aligned}
    &S^{J}_{N\tau nl,N'\tau'n'l'} =
    \sum\limits_{\Omega}
    \sum\limits_{\lambda\lambda'}
    \sum\limits_{KK'}
    (-1)^{l+l'+\lambda+\lambda'}
    \\& \times
    c^{(N\tau)*}_{K}
    c^{(N'\tau')}_{K'}
    C^{NK}_{l-\lambda,J\Omega}
    C^{N'K'}_{l'-\lambda',J\Omega}
    S_{nl\lambda,n'l'\lambda'},
  \end{aligned}
  \label{eqn:app:RFT2}
\end{equation}
using standard Clebsch-Gordan permutation relations.

\bibliography{refs}

\end{document}